\documentclass[a4paper,fleqn,usenatbib]{mnras}

\usepackage{newtxtext,newtxmath}
\usepackage[T1]{fontenc}
\usepackage{ae,aecompl}

\usepackage{graphicx}	% Including figure files
\usepackage{amsmath}	% Advanced maths commands
\usepackage{sidecap}

\title[Satellites of Milky Way analogues]{Can cosmological simulations capture the diverse satellite populations of observed Milky Way analogues?}
\author[A. S. Font et al.]{
  Andreea S. Font,$^{1}$\thanks{E-mail: A.S.Font@ljmu.ac.uk}, Ian G. McCarthy,$^1$ Vasily Belokurov$^2$\\
$^{1}$Astrophysics Research Institute, Liverpool John Moores University, 146 Brownlow Hill, Liverpool L53RF, UK\\
$^{2}$Institute of Astronomy, University of Cambridge, Madingley Road, Cambridge CB3 0HA, UK
}

\date{Accepted XXX. Received YYY; in original form ZZZ}

\pubyear{2020}

\begin{document}
\label{firstpage}
\pagerange{\pageref{firstpage}--\pageref{lastpage}}
\maketitle

\begin{abstract}
The recent advent of deep observational surveys of local Milky Way `analogues' and their satellite populations allows us to place the Milky Way in a broader cosmological context and to test models of galaxy formation on small scales.  In the present study, we use the $\Lambda$CDM-based ARTEMIS suite of cosmological hydrodynamical simulations containing 45 Milky Way analogue host haloes to make comparisons to the observed satellite luminosity functions, radial distribution functions, and abundance scaling relations from the recent Local Volume and SAGA observational surveys, in addition to the Milky Way and M31.  We find that, contrary to some previous claims, $\Lambda$CDM-based simulations can successfully and simultaneously capture the mean trends and the diversity in both the observed luminosity and radial distribution functions of Milky Way analogues once important observational selection criteria are factored in.  Furthermore, we show that, at fixed halo mass, the concentration of the simulated satellite radial distribution is partly set by that of the underlying smooth dark matter halo, although stochasticity due to the finite number of satellites is the dominant driver of scatter in the radial distribution of satellites at fixed halo mass.
\end{abstract}


\section{Introduction}

The formation of the Milky Way has traditionally been considered as a blueprint for understanding the formation of normal spiral galaxies in general.  The adoption of this assumption has perhaps been mainly motivated by the fact that, typically, there has been much higher quality and detailed observations available for the Milky Way than for any other galaxy.  To what extent the Milky Way is actually representative of disc galaxies is presently uncertain, though.  With the advent of new deep, dedicated extragalactic surveys of so-called ``Milky Way analogues'' (i.e., disc galaxies with total halo masses of $\approx 10^{12} \, {\rm M}_{\odot}$) in the Local Volume \citep{danieli2017,smercina2018,bennet2019,crnojevic2019,bennet2020,carlsten2020_survey} and out to $\sim 40$~Mpc \citep{geha2017,mao2021}, there is a rapidly diminishing requirement to rely on the Milky Way as our template for disc galaxy formation.  Instead, the rapid increase in the number of Milky Way-mass galaxies with high-quality data available, both in observations and in cosmological simulations, motivates a reassessment of what constitutes a typical disc galaxy and how the Milky Way fits into this picture.

Work along these lines has suggested that the Milky Way may not have had a typical merger history for a galaxy of total mass of $\approx 10^{12} \, {\rm M}_{\odot}$.  In particular, the Milky Way appears to have had a very quiescent history since $z \simeq 2$ (e.g., \citealt{wyse2001,Deason2013,ruchti2015,Lancaster2019}), which is relatively rare in a cosmological context (e.g., \citealt{stewart2008,font2017}).  In contrast, M31, another well-studied disc galaxy of similar mass to our own, shows evidence of a much more active accretion history, as indicated by its disturbed stellar disc, ubiquitous tidal debris in its stellar halo and a significantly higher abundance of satellite galaxies (e.g. \citealt{mcconnachie2009,mcconnachie2018,Deason2013,dSouza2018}).  As galaxies and their dark matter haloes in $\Lambda$CDM are, to a large extent, assembled through the accretion and mergers of smaller satellite galaxies, the properties of the satellite population (e.g., its abundance, spatial distribution, internal properties) contain a significant amount of information about the formation histories of their hosts.  They are effectively proxies for the formation histories of galaxies and by examining the satellite populations\footnote{A caveat, of course, is that some satellites are completely disrupted and/or merged with the central galaxy, implying that the current satellite population does not contain a complete census of merger history of the system.  But this information is ultimately retained in the properties of the stellar halo and the central galaxy and is therefore potentially recoverable.} of hosts of (approximately) fixed mass, we are effectively examining the differing formation histories that lead to galaxies of a fixed mass.  

Advances in observational surveys have resulted in growing samples of dwarf satellite galaxies around dozens of Milky Way analogues.  Given the distances to these galaxies, the samples are complete only in the `classical' dwarf regime (e.g. limiting magnitudes of $M_V \simeq -8$ or $-9$ for galaxies in the Local Volume or to $M_r \simeq -12$ out to $20-40$~Mpc). Nevertheless, the properties of Local Group classical dwarf galaxies can now be put into a broader `cosmological' context. 

Apart from the question of Milky Way's typicality, there is also the question of whether theoretical models (specifically, whether cosmological simulations based on a $\Lambda$CDM cosmology) can generally reproduce the properties of observed satellite galaxies.  The current cosmological model has been confronted with several `small-scale problems', including: the so-called missing satellite problem (e.g., \citealt{kauffmann1993,klypin1999,moore1999}), the `too big to fail' problem \citep{read2006,boylan-kolchin2011,boylan-kolchin2012}, and the cusp-core problem \citep{flores1994,moore1994}. These apparent problems are present in both the Milky Way and M31  \citep{tollerud2014}.  A variety of potential solutions to these problems have been proposed (both within the context of $\Lambda$CDM and beyond), starting with observational bias corrections \citep[e.g.][]{Koposov2008, Tollerud2008, Koposov2009, Jethwa2018, Kim2018} and invoking baryonic processes \citep{governato2012,pontzen2012,brooks2014,sawala2016,wetzel2016}, changing the nature of dark matter, such as warm or self-interacting dark matter \citep{spergel2000,rocha2013,lovell2014}, changing the nature of gravity (e.g., \citealt{brada2000}), or the values of some of the cosmological parameters that preferentially affect small scales, such as the running of the scalar spectral index of primordial fluctuations \citep{garrison-kimmel2014,stafford2020}.  In the present study it is not our aim to re-examine these well-studied questions.  

The observations of Milky Way analogues and of their satellite populations provide a renewed motivation to test the predictions of cosmological models. Although most of the small-scale problems above require resolved stellar spectroscopy, which is not currently possible beyond the Local Group, observations of distant Milky Way analogues provide other useful small-scale tests. In addition to determining the satellite luminosity functions, they also provide measurements of the largest magnitude gaps in these functions (which can give an indication of their slopes) and of the spatial distributions of dwarf galaxies. With a growing sample of Milky Way-mass galaxies, these observables can in principle be correlated with the properties of host galaxies. Additionally, the system-to-system {\it scatter} in the observables (e.g., luminosity functions and radial distributions) can be quantified and compared with theoretical predictions.  Needless to say, a successful theory should not only capture the mean trends but also the scatter about them.

Observations of Milky Way analogues to date have already revealed some new and interesting puzzles about the populations of classical dwarfs. These include:

\begin{itemize}
    \item {\it Too large scatter in the luminosity functions?} Some disc galaxies display a strikingly low number of bright satellites compared with the Milky Way.  For example, M94, dubbed as the `lonely giant' \citep{smercina2018} has only two satellites brighter than $M_V\sim -9$ within $\approx 150$~kpc. Another large disc galaxy, M101, has only $9$  satellites brighter than $M_V \sim -8$ within the same radius \citep{danieli2017,bennet2019,bennet2020}.  For comparison, the median number of classical dwarf satellites per host in the Local Volume is $\approx 24$ \citep{bennet2019}, a number that is similar to that found for M31 \citep{mcconnachie2012}. Cosmological hydrodynamical simulations that contain a statistical sample of Milky Way-mass galaxies with their satellite populations can potentially elucidate whether sparse systems such as M94 or M101 occur naturally in a $\Lambda$CDM model.
    
    \item {\it Tensions between the observed and predicted radial distributions of satellites}. The radial distribution of Milky Way satellites appears to stand in contrast with the predictions of $\Lambda$CDM models, in the sense that it appears to be more centrally-concentrated than typical simulated Milky Way-mass systems \citep{kravtsov2004,yniguez2014,samuel2020}.  Previous studies have shown that the incorporation of important baryonic physics can help to reconcile this tension (e.g., \citealt{kravtsov2004,maccio2010,font2011,starkenburg2013}).  Alternatively, or in addition to, it is possible that simulations with limited numerical resolution may suffer from the spurious tidal disruption of satellites (e.g., \citealt{vandenbosch2018}) or that an overly energetic feedback results in satellites being centrally cored and therefore more vulnerable to disruption. Another possibility is that the discrepant radial distributions may be reconciled if the observed tally of dwarf galaxies beyond $\sim 100$~kpc from the centre of the Milky Way is incomplete \citep{garrison-kimmel2019}. As we will discuss in this paper, we find an inverse radial distribution problem, where dwarf galaxies around some Milky Way analogues in the SAGA survey \citep{mao2021} appear, at face value, to be significantly less centrally-concentrated than predicted.  Therefore, careful comparisons between models and observations are required in order to understand the causes of these apparently contradictory results.
    
    \item As already noted, a number of possible scaling relations between the abundance of surviving satellites and the properties of their hosts galaxies have been examined, including correlations with the host total mass \citep{trentham2009,starkenburg2013,fattahi2016,garrison-kimmel2019}, stellar mass or magnitude \citep{geha2017}, or even with the bulge-to-total ratio of the host \citep{javanmardi2020}.  The local environment in which the host galaxy lives is also believed to play a role and a relation between the number of satellites and tidal index has ($\Theta_{5}$) been investigated \citep{karachentsev2013,bennet2019}. If confirmed, these relations can help to further constrain the assembly histories of galaxies.  We will examine to what extent cosmological simulations faithfully capture these correlations and whether/how issues such as selection effects limit observational analyses. 
    
\end{itemize}

This study uses a new suite of zoomed-in, cosmological hydrodynamical simulations of Milky Way-mass galaxies called ARTEMIS (Assembly of high-ResoluTion Eagle-simulations of MIlky Way-type galaxieS).  The suite 
comprises $45$ such systems and their retinue of dwarf galaxies.  The simulations have previously been shown to match a range of global properties of Milky Way-mass galaxies, such as galaxy sizes, star formation rates, stellar metallicities, and various observed properties of Milky Way-mass stellar haloes \citep{font2020} and of the solar neighborhood \citep{poole-mckenzie2020}. Here we focus on the properties of simulated satellite galaxies and we make comparisons with observations of Milky Way analogues, as described above.

The paper is structured as follows. In Section~\ref{sec:data} we briefly describe the ARTEMIS simulations and the observational samples used in this study. In Section~\ref{sec:results} we analyse the luminosity functions of the simulated galaxies, the abundance of satellites in relation to various properties of their hosts, and the radial distributions of satellites. Throughout this study, we compare our results with observations in the Milky Way, M31 and other Milky Way analogues. In Section~\ref{sec:concl} we summarise our findings and conclude.

\section{Simulations and observational data sets}
\label{sec:data}

\subsection{The ARTEMIS simulations}
\label{sec:sims}

The ARTEMIS suite comprises $45$ zoomed hydrodynamical simulations of Milky Way-mass haloes.  The majority ($42$) of these systems were introduced in \citet{font2020}, to which we add $3$ new systems constructed with the same methods.  As described in \citet{font2020}, these systems were selected from a periodic box of $25$~Mpc/$h$ on a side.  The selection criterion was based solely on halo mass, specifically, that their total mass in the parent dark matter-only periodic volume be in the range of $8\times10^{11} < {\rm M}_{200}/{\rm M}_\odot < 2\times10^{12}$, where ${\rm M}_{200}$ is the mass enclosing a mean density of 200 times the critical density at $z=0$. The $45$ systems were re-simulated at higher resolution with hydrodynamics in a $\Lambda$CDM cosmological model. The cosmological parameters correspond to the maximum likelhood WMAP9 $\Lambda$CDM cosmology with $\Omega_m=0.2793$, $\Omega_b=0.0463$, $h=0.70$, $\sigma_8=0.8211$, $n_s=0.972$ \citep{hinshaw2013}. 

In the zoom simulations, the initial baryonic particle mass is $2.23 \times 10^4 \, {\rm M}_{\odot}/h$, the dark matter particle mass is $1.17 \times 10^5 \, {\rm M}_{\odot}/h$ and the force resolution (the Plummer-equivalent softening) is $125$ pc/$h$. The simulations were run with a version of the Gadget-3 code with galaxy formation (subgrid) physics models developed for the EAGLE simulations \citep{schaye2015}. These include prescriptions for metal-dependent radiative cooling in the presence of a photo-ionizing UV background, star formation, supernova and active galactic nuclei feedback, stellar and chemical evolution, formation of black holes (for details of these physical prescriptions see \citealt{schaye2015} and references therein).  In terms of the galaxy formation modelling, the main difference between ARTEMIS and EAGLE relates to the stellar feedback scheme, which was adjusted in ARTEMIS to achieve an improved match to the amplitude of the stellar mass--halo mass relation as inferred from recent empirical models \citep{moster2018,behroozi2019}.  As discussed in \citet{font2020}, this was achieved in practice by increasing the density scale where the energy used for stellar feedback becomes maximal\footnote{The fraction of available stellar feedback energy used for feedback is modelled with a sigmoid function of density (and metallicity) in the EAGLE code.  The sigmoid function asymptotes to fixed values at low and high densities, such that a higher fraction of the available energy is used at high densities in order to offset spurious (numerical) radiative cooling losses.  As we increase the resolution of the simulations, the density scale at which numerical losses become important increases, motivating an increase in the transition density scale used for stellar feedback.}.  As shown in \citet{font2020}, the simulations not only reproduce the stellar mass of Milky Way-mass haloes (which is by construction), but they also reproduce the observed sizes and star formation rates of such systems but without any explicit calibration to match those quantities.

In order to make more meaningful comparisons with optical observations, we compute in post-processing the luminosities/magnitudes of star particles in various bands by assuming the star particles are simple stellar populations (SSPs).  Given the age, metallicity and initial stellar mass of each star particle, we use the PARSEC v1.2S+COLIBRI PR16 isochrones \citep{bressan2012,marigo2017} to compute dust-free luminosities and magnitudes.  In doing so, we adopt the same \citet{chabrier2003} stellar initial mass function used in the simulations.

As noted above, no constraints were imposed on the merger histories of the simulated Milky Way-mass haloes. This choice stems from our aim to capture the diversity of formation scenarios for galaxies of this mass. The majority of simulated hosts have a disc morphology at $z = 0$, supporting the findings of previous studies that disc galaxies can form under a variety of merger scenarios \citep{font2017}. For example, the co-rotation parameter, $\kappa_{co}$, which measures the (mass-weighted) fraction of kinetic energy in ordered rotation, ranges between $0.2 - 0.8$, with a typical value of $\simeq 0.4$ (see \citealt{font2020} for details). Other properties of the simulated Milky Way-mass galaxies can be found in Table~1 of \citet{font2020}.  

The focus of the present study is on comparing the predicted and observed properties of satellite populations around Milky Way-mass haloes.  It is therefore important that we define what a satellite is, in order to make a consistent comparison between the simulations and observations.  Typically, for simulation-based studies, satellites correspond to those subhaloes (here identified with \textsc{SUBFIND}; \citealt{dolag2009}) which are gravitationally bound to their host and, typically, located within some radius, such as ${\rm R}_{200}$, centred on the host.  

In observations, the adopted definitions for what constitutes a satellite can differ from study to study. This is primarily because one cannot accurately determine the size or the total mass of the host galaxy or measure 3D distances between potential satellites and the host.  Therefore, rather than adopt the standard physical `simulator' definition, for comparisons with the results of specific surveys we will adopt the observational selection criteria, which in some cases means we will include subhaloes/galaxies that are outside ${\rm R}_{200}$ (e.g., within a fixed distance of $300$~kpc) and that are potentially unbound as well.  These simulated systems will be referred to as dwarf galaxies. For observed systems we retain the term `satellite galaxy' used in the original studies, irrespective of whether they are truly satellites of their host galaxy.

Finally, in the simulations we limit our analysis to satellites with a minimum of $10$ star particles, corresponding to stellar masses $\ga 2 \times 10^5 \, {\rm M}_{\odot}/h$ or, roughly, $M_V<-8$.  Therefore, we focus on the regime of classical dwarf galaxies observed in the Milky Way and in the Milky Way analogues.  Note that for satellites near this low particle number threshold we expect finite resolution to affect their internal properties (e.g., sizes).  However, we can reliably measure certain global properties\footnote{We use the phrase `reliably measure' in the sense that we can reliably derive those quantities based on the particles bound to the subhalo.  This is not the same as saying those properties are robust to large changes in numerical resolution.  In order to determine that we would require even higher resolution simulations, which are prohibitively expensive.}, specifically their integrated stellar masses and luminosities (within Poisson uncertainties) and distances from their host.

\begin{figure}
    \includegraphics[width=\columnwidth]{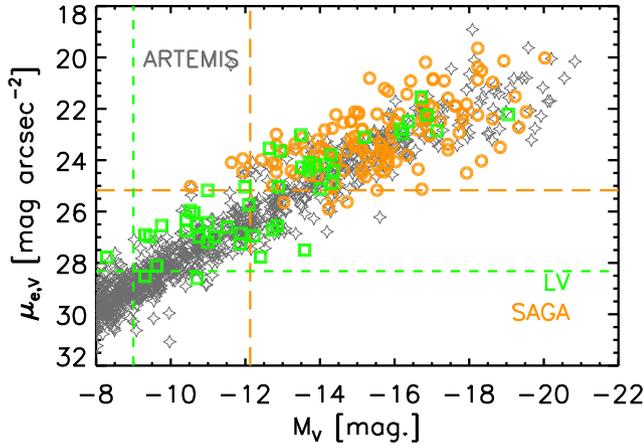}
    \caption{The $V$-band magnitude--effective surface brightness relations of satellites of Milky Way-mass host galaxies in ARTEMIS (black diamonds), SAGA (orange circles), and the Local Volume (green squares).  For SAGA we have converted the published r-band magnitudes to the V-band assuming a colour $r-V = -0.17$ (the median value from the LV sample).  The dashed vertical lines correspond to quoted observational magnitude limits.  For the dashed green horizontal line we have converted the quoted central surface brightness limit of $\mu_0,V = 26.5$ mags.~arcsec$^{-2}$ from \citet{carlsten2021} into an effective surface brightness limit assuming an expontential light profile ($28.3$ mags.~arcsec$^{-2}$).  The dashed orange horizontal line corresponds to $\mu_{{\rm eff},r} = 25$ mags. arcsec$^{-2}$ which the SAGA data appear to obey.  We impose these respective magnitude and surface brightness cuts to the ARTEMIS simulations when comparing to the SAGA and LV surveys.}
    \label{fig:mueff_mag}
\end{figure}

\subsection{Observational samples}
\label{sec:obs}

Throughout this study we will compare the simulated Milky Way-mass galaxies with observations in the Local Group, namely with the Milky Way and M31, and with Milky Way analogues in the Local Volume and out to larger distances of $\sim 20-40$~Mpc. 

For the Local Volume, we use data from \citet{carlsten2020_survey,carlsten2020_radial,carlsten2021} and references therein. However, we only use hosts of approximately Milky Way-mass, specifically with estimated total masses of $(1-3) \times 10^{12} \, {\rm M}_{\odot}$, and with a complete sample of classical satellites within $\sim 150$~kpc from the center of their host. This selection yields 6 Milky Way analogues: NGC 4565, NGC 4631, NGC 4258, M51, M94 and M101. These satellite samples are complete typically down to $M_V \approx -9$, with the exception of NGC 4565, where the completeness extends down to $M_V \approx -12$. 
 
For Milky Way analogues beyond $20$~Mpc, we use the data from the SAGA survey, in particular the recent `stage 2' release \citep{mao2021}. In this survey, all dwarfs within a projected distance of $300$~kpc and within a velocity of $\pm 250$~km s$^{-1}$ from their hosts are considered to be satellites. This choice is motivated by the fact that, in a $\Lambda$CDM cosmology, a galaxy with virial mass of $1.6 \times 10^{12} \, {\rm M}_{\odot}$ (or ${\rm M}_{200} \approx 1.4 \times 10^{12} \, {\rm M}_{\odot}$) has a virial radius of approximately $300$~kpc (or ${\rm R}_{200} \approx 220$ kpc).  This SAGA stage 2 sample contains $34$ Milky Way analogues. 

Given the different selection criteria of these two observational data sets, we choose to compare the simulations with each set individually. This allows us to tailor the comparison to the specifics of these surveys. For example, for comparisons with galaxies in the Local Volume, we select only simulated satellites with $M_V<-9$ and within $150$~kpc of their hosts, while for comparisons with SAGA we select all dwarf galaxies with $M_r < -12.3$ within a projected distance of $300$~kpc.   For physical interpretations, we employ satellites within more physically motivated parameters, e.g. ${\rm R}_{200}$. 

An important question is whether or not the adopted spatial and magnitude cuts are sufficient to enable a fair comparison between the simulations and the observational surveys. Generally speaking, we expect galaxy selection to be influenced not only by magnitude but also surface brightness (e.g., a relatively luminous galaxy that is very spatially extended can have a low surface brightness and possibly avoid selection).  Therefore, in Fig.~\ref{fig:mueff_mag} we show the magnitude--effective surface brightness relations of satellite galaxies in ARTEMIS (black diamonds), SAGA (orange circles), and the Local Volume (green squares).  \citet{carlsten2021} estimate a central surface brightness limit for the Local Volume survey of $\mu_{0,V} \approx 26.5$ mags. arcsec$^{-2}$.  We convert this into an effective surface brightness limit (i.e., within the projected half-light radius) assuming an exponential light profile, leading to a reduction of 1.822 \citep{graham1997}, or $\mu_{{\rm eff},V} \approx 28.3$ mags. arcsec$^{-2}$.  No surface brightness limit is explicitly quoted for the SAGA survey, but the effective surface brightnesses (also assuming an exponential profile) are provided for the selected satellites in \citet{mao2021}.  To directly compare the magnitudes and surface brightnesses of the Local Volume and SAGA surveys, we apply a simple fixed colour correction of $r-V = -0.17$ to the SAGA survey.  

From Fig.~\ref{fig:mueff_mag} we can see that, as indicated in \citet{carlsten2021}, the Local Volume satellites extend down to $\mu_{{\rm eff},V} \approx 28.3$ mags. arcsec$^{-2}$ in their deep CFHT MegaCam imaging.  The SAGA survey, for which satellites are initially identified in shallower SDSS (stage 1) and DECaLS and DES (stage 2) observations, does not identify and/or follow up many satellites below $\approx 25$ mags. arcsec$^{-2}$.  Comparing the Local Volume and SAGA relations, the SAGA survey hits this apparent surface brightness limit before hitting the quoted magnitude limit.  That is, according to the Local Volume survey, there do exist satellites brighter than the SAGA magnitude limit but which appear to be selected against in SAGA due to their relatively low surface brightness.  (The strong tapering in the SAGA relation at low luminosities also suggests this.)  This motivates us to include surface brightness cuts in the comparisons to the Local Volume and SAGA surveys.  Specifically, we apply cuts of $\mu_{{\rm eff},V} = 28.3$ mags. arcsec$^{-2}$ and $\mu_{{\rm eff},r} = 25.0$ mags. arcsec$^{-2}$ when comparing to the Local Volume and SAGA surveys, respectively.  These are in addition to the magnitude and spatial cuts described above.  Collectively, we refer to the combined spatial, magnitude, and surface brightness criteria as either `LV selection' or `SAGA selection' when discussing the simulations.  We will comment on the importance of the surface brightness limits where relevant.  When comparing to the satellite systems of the Milky Way and M31 (which we do separately from comparisons to the Local Volume and SAGA), we do not expect surface brightness limits to be an issue.  We adopt a magnitude cut of $M_V = -8$ and a maximum radius of 300 kpc in this case.

We note that, at least qualitatively speaking, the ARTEMIS simulations reproduce the observed magnitude--surface brightness relations above these limits which is a non-trivial result, particularly as the simulations were not used to inform the observation limits in any way.

\section{Results}
\label{sec:results}

In this section we present the main results of our study.  We first discuss the properties of the host galaxies (Section \ref{sec:hosts}), we then examine the satellite luminosity functions (Section \ref{sec:lumfunc}), the relations between the abundance of satellites and various host properties (Section \ref{sec:abundance}), and the radial distribution of satellites (Section \ref{sec:radial}).

\subsection{Properties of the hosts}
\label{sec:hosts}

In Table~\ref{tab:table1} in Appendix \ref{sec:appendixA} we list a number of properties of the $45$ Milky Way-mass galaxies: their total spherical-overdensity masses (${\rm M}_{200}$ and virial masses, ${\rm M}_{\rm vir}$, both defined with respect to the critical density of the universe), the corresponding radii (${\rm R}_{200}$ and ${\rm R}_{\rm vir}$) and magnitudes of the central galaxy in various bands ($K$, $B$, $V$ [Vega] and the SDSS $r$-band [AB]).

\begin{figure}
    \includegraphics[width=\columnwidth]{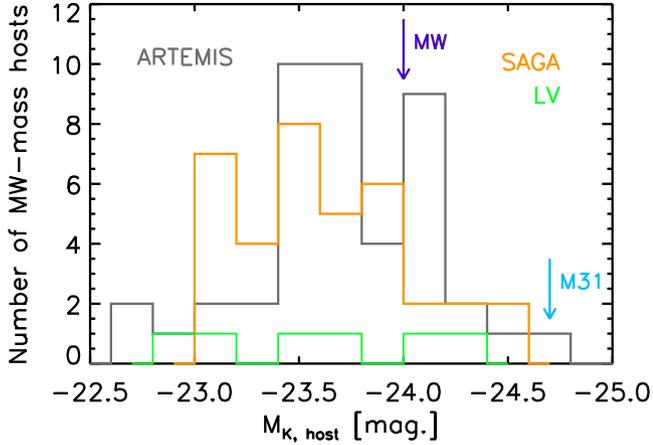}
    \caption{The distribution of $K$-band magnitudes ($M_K$) of the host galaxies in ARTEMIS (black), compared with the observational samples used in this study: Milky Way analogues in SAGA (orange) and in the Local Volume (green).  The vertical arrows indicate the value for the Milky Way and M31.}
    \label{fig:hist_mag}
\end{figure}

Fig.~\ref{fig:hist_mag} shows the distribution of $K$-band magnitudes, which is a good observational proxy for stellar mass, of the hosts in ARTEMIS in comparison with those of Milky Way analogues in the Local Volume and SAGA surveys.  We use the K-band magnitudes quoted in the Local Volume and SAGA studies, apart from M94 and M101 (LV) for which no estimates of the K-band magnitudes were provided.  For these two galaxies we use the apparent K-band magnitudes from the 2MASS survey\footnote{https://irsa.ipac.caltech.edu/Missions/2mass.html}, which are 5.106 and 5.512 (respectively), with the distances given in \citet{carlsten2021} to estimate the absolute magnitudes for M94 and M101: $-23.01$ and $-23.56$, respectively.
The $M_K$ distributions of the three samples are similar.  The median (mean) magnitudes are: $-23.69$ ($-23.84$) for ARTEMIS, $-23.57$ ($-23.72$) for SAGA, and $-23.80$ ($-23.75$) for the Local Volume.  Note that we compute the mean magnitudes as the magnitude corresponding to the mean luminosity, rather than (incorrectly) averaging magnitudes.  The median (mean) K-band luminosities (in solar units) are: $6.07\times10^{10}$ ($7.00\times10^{10}$) for ARTEMIS, $5.45\times10^{10}$ ($6.23\times10^{10}$) for SAGA and $6.73\times10^{10}$ ($6.45\times10^{10}$) for the Local Volume.  Thus, the median and mean luminosities agree between the three datasets to approximately 12\% accuracy for the two most disparate datasets (ARTEMIS and SAGA).  As we show later (see Fig.~\ref{fig:Nsat_scaling}), the abundance of satellites does not depend particularly strongly on the K-band magnitude when observational selection effects are included, implying that the slight differences in the K-band magnitude distributions of the three datasets will not affect our conclusions.  Nevertheless, we will comment throughout on the impact of scaling out the dependence on host magnitude.

Note that the median values of these sets are slightly below (less bright than) the Milky Way's value ($M_K \approx -24$; \citealt{drimmel2001}. M31 has $M_K \approx -24.7$; \citealt{geha2017}), but there are several Milky Way analogues closely matching the Milky Way itself. 

The $V$-band magnitudes of the simulated galaxies (presented in Table~\ref{tab:table1}) also agree reasonably well with the observed values. For example, the range in the simulations is from $-19.82$ to $-21.96$. The Milky Way value is $M_V=-21.37$ \citep{bland-hawthorn2016}, while for other Milky Way analogues, $M_V$ ranges between $\simeq -19.95$ for M94 \citep{gildepaz2007} to $\simeq -22$ for M31 \citep{walterbos1987}. 

The range in the virial masses for the simulated galaxies is $(0.76 - 4.21) \times 10^{12} \, {\rm M}_{\odot}$, with a median value of $1.15 \times 10^{12} \, {\rm M}_{\odot}$ (the median ${\rm M}_{200}$ value is $\approx 1 \times 10^{12} \, {\rm M}_{\odot}$). This range covers the various estimates of the virial masses of the Milky Way and M31, including their uncertainties. Previously, for the Milky Way, total mass estimates as high as $1.67 \times 10^{12} \, {\rm M}_{\odot}$ \citep{wilkinson1999} or even $2.62 \times 10^{12} \, {\rm M}_{\odot}$ \citep{watkins2010} were produced. On the other hand, numbers as low as $0.55 \times 10^{12} \, {\rm M}_{\odot}$ \citep{gibbons2014} were also reported.  Very recently, however, there appears to be some convergence in determinations of the Milky Way mass using {\it Gaia} data \citep[see e.g.][]{Callingham2019, Posti2019, Watkins2019,Deason2019,Vasiliev2019,Eadie2019,Erkal2019,cautun2020}, with different methods reporting a value close to $1.2 \times 10^{12} \, {\rm M}_{\odot}$ (for the total mass of the Galaxy including the baryonic component and the Magellanic Cloud).  The uncertainties in this figure vary from study to study, depending on the prior information assumed about the concentration of the halo and/or the velocity anisotropy of the employed tracers, but is typically $20-40\%$.  For M31, the estimates range from $1.7 \times 10^{12} \, {\rm M}_{\odot}$ \citep{diaz2014} at the high end down to $0.8 \times 10^{12} \, {\rm M}_{\odot}$ \citep{kafle2018}.  The other Milky Way analogues used in this study also have estimated virial masses within the range $(1-3) \times 10^{12} \, {\rm M}_{\odot}$ (see \citealt{karachentsev2014} for galaxies in the Local Volume and \citealt{geha2017} for those in the SAGA survey).

Generally speaking, therefore, there is significant overlap in the halo masses and luminosities of the simulated and observed host systems, allowing for a detailed comparison of their satellite properties (below).  It is worth highlighting here that while the stellar feedback in ARTEMIS was calibrated to reproduce the empirical stellar mass--halo mass relation at the Milky Way mass scale (and thus we expect the host luminosities to be realistic by construction for Milky Way-halo mass systems), the properties of satellite galaxies were not examined at any part of this process.  Thus, they represent an interesting and potentially challenging test for the simulations.

\subsection{Luminosity functions}
\label{sec:lumfunc}

Having established that the host systems of the simulated and observational samples are similar, we now examine the satellite luminosity functions of those hosts.  We begin by comparing to the Milky and M31 before turning our attention to comparisons with the Local Volume and SAGA surveys.

\begin{figure}
    \includegraphics[width=\columnwidth]{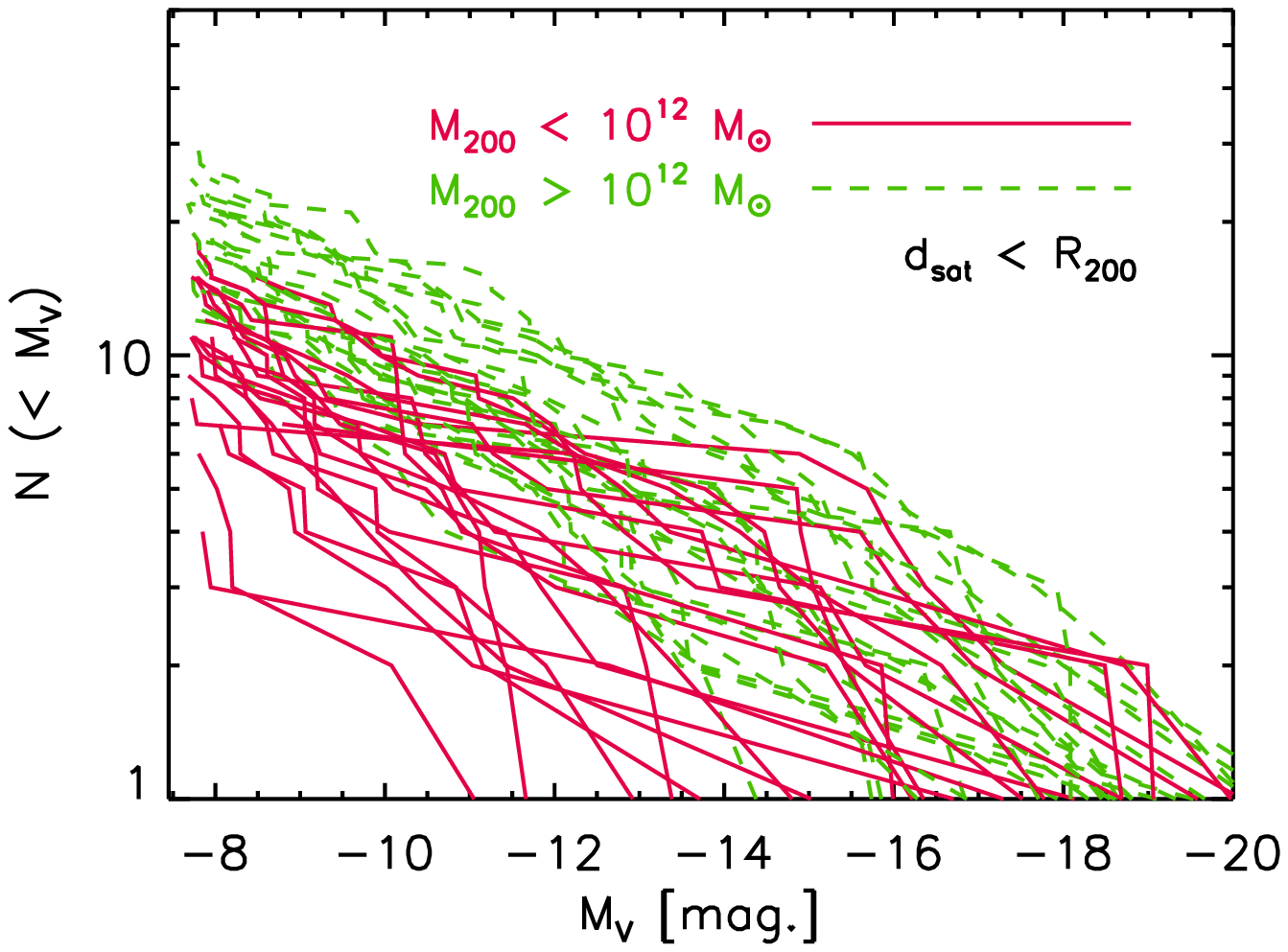}
    \includegraphics[width=\columnwidth]{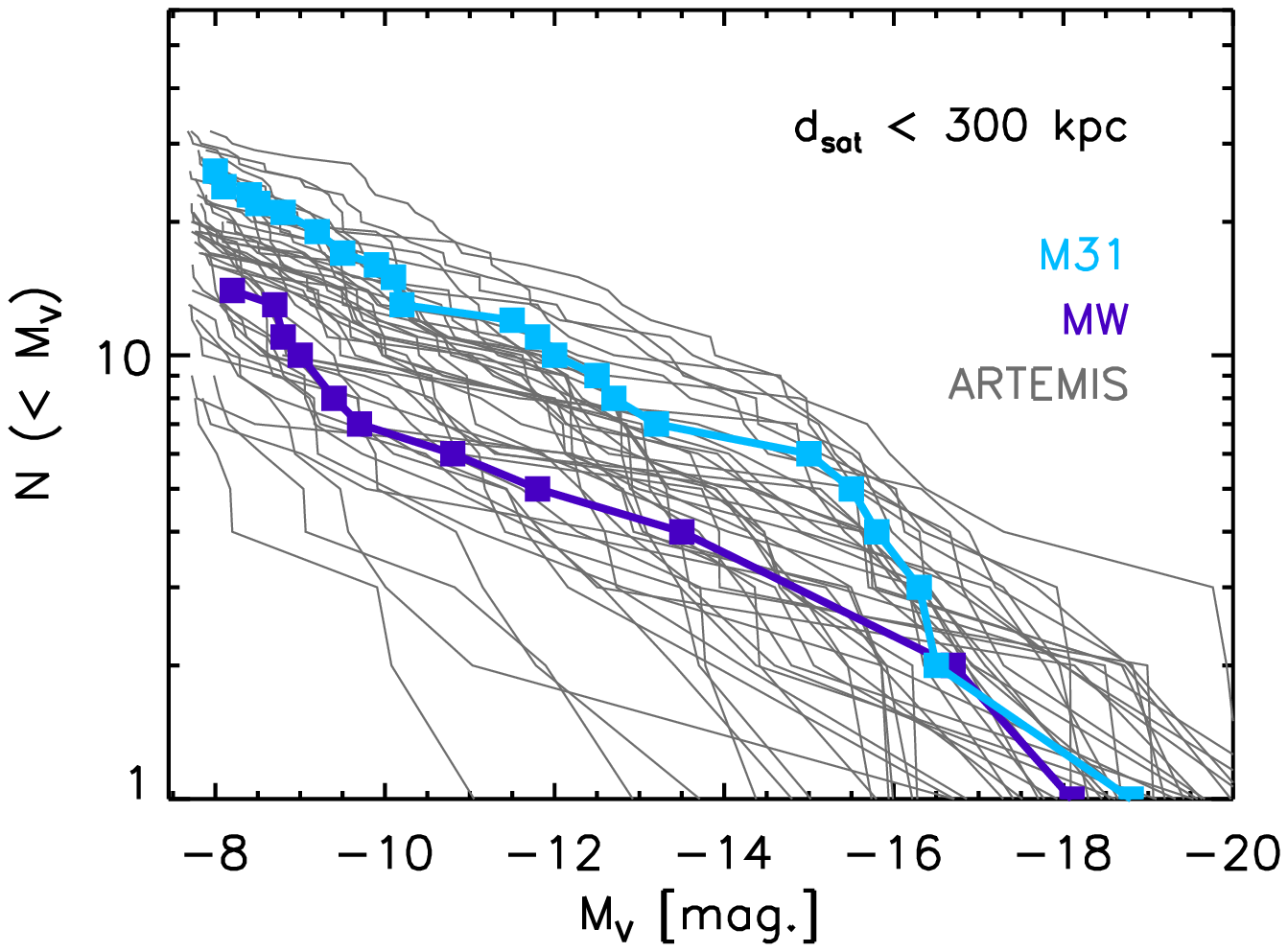}
    \includegraphics[width=\columnwidth]{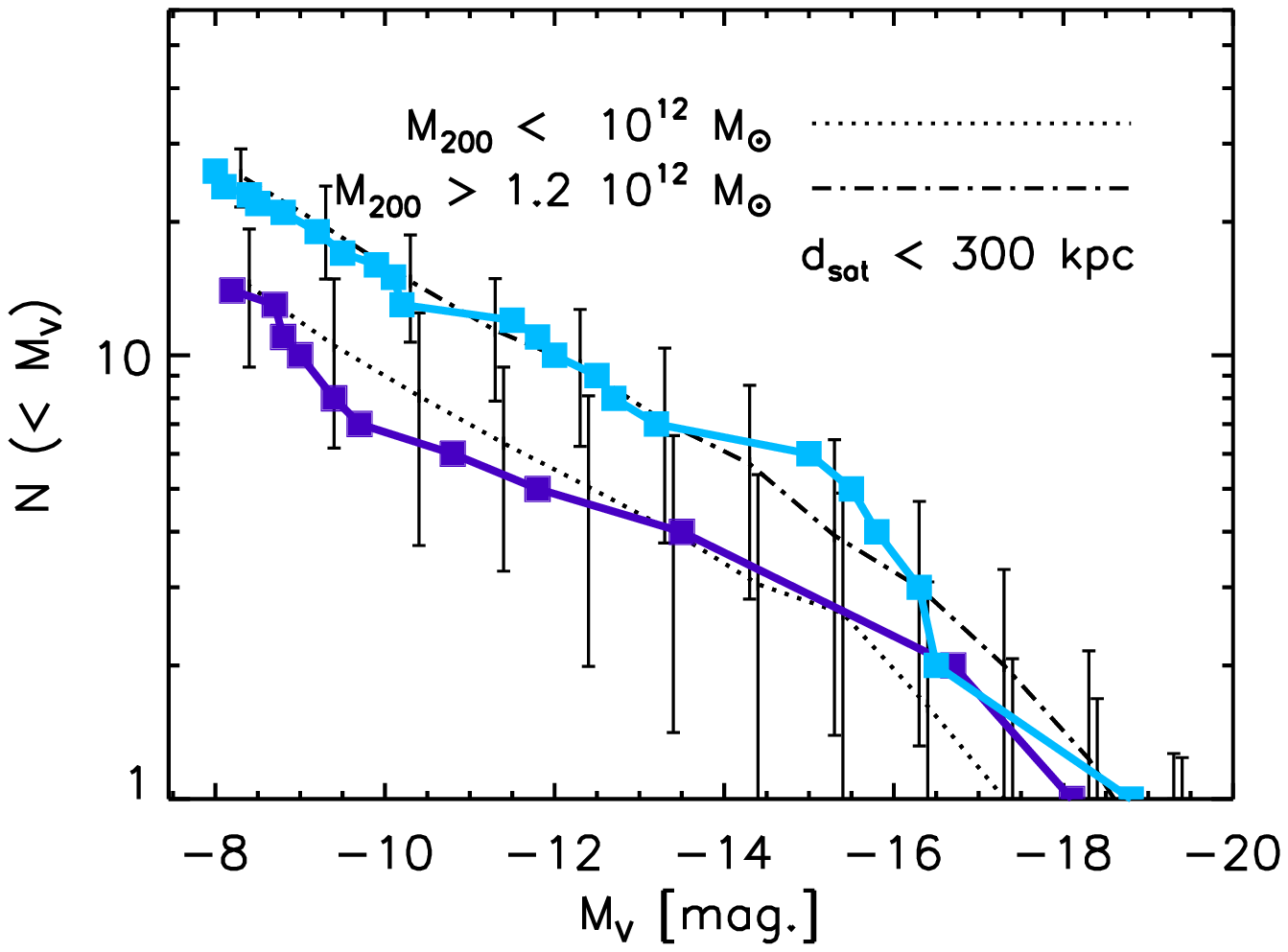}
    \caption{Satellite luminosity functions. {\it Top:} $V$-band luminosity functions of satellites within ${\rm R}_{200}$ in ARTEMIS. Massive systems (${\rm M}_{200} > 10^{12} \, {\rm M}_{\odot}$) are shown with dashed lines, while less massive systems are shown with full lines. {\it Middle}: Luminosity functions of dwarf galaxies within $300$~kpc in simulations (grey lines) compared with those in the Milky Way (dark blue) and M31 (cyan). {\it Bottom:} Similar comparison as above, but showing which simulated hosts match each galaxy. Lower mass systems (${\rm M}_{200}< 10^{12}\, {\rm M}_{\odot}$) match better the Milky Way, while the higher mass ones (${\rm M}_{200} >1.2 \times 10^{12}\, {\rm M}_{\odot}$). The two ${\rm M}_{200}$ median values are $7.9 \times 10^{11} \, {\rm M}_{\odot}$ and $1.6 \times 10^{12} \, {\rm M}_{\odot}$, respectively.}
    \label{fig:lfunc_MW_M31}
\end{figure}

\begin{figure*}
  \begin{tabular}{cc}
  \includegraphics[width=\columnwidth]{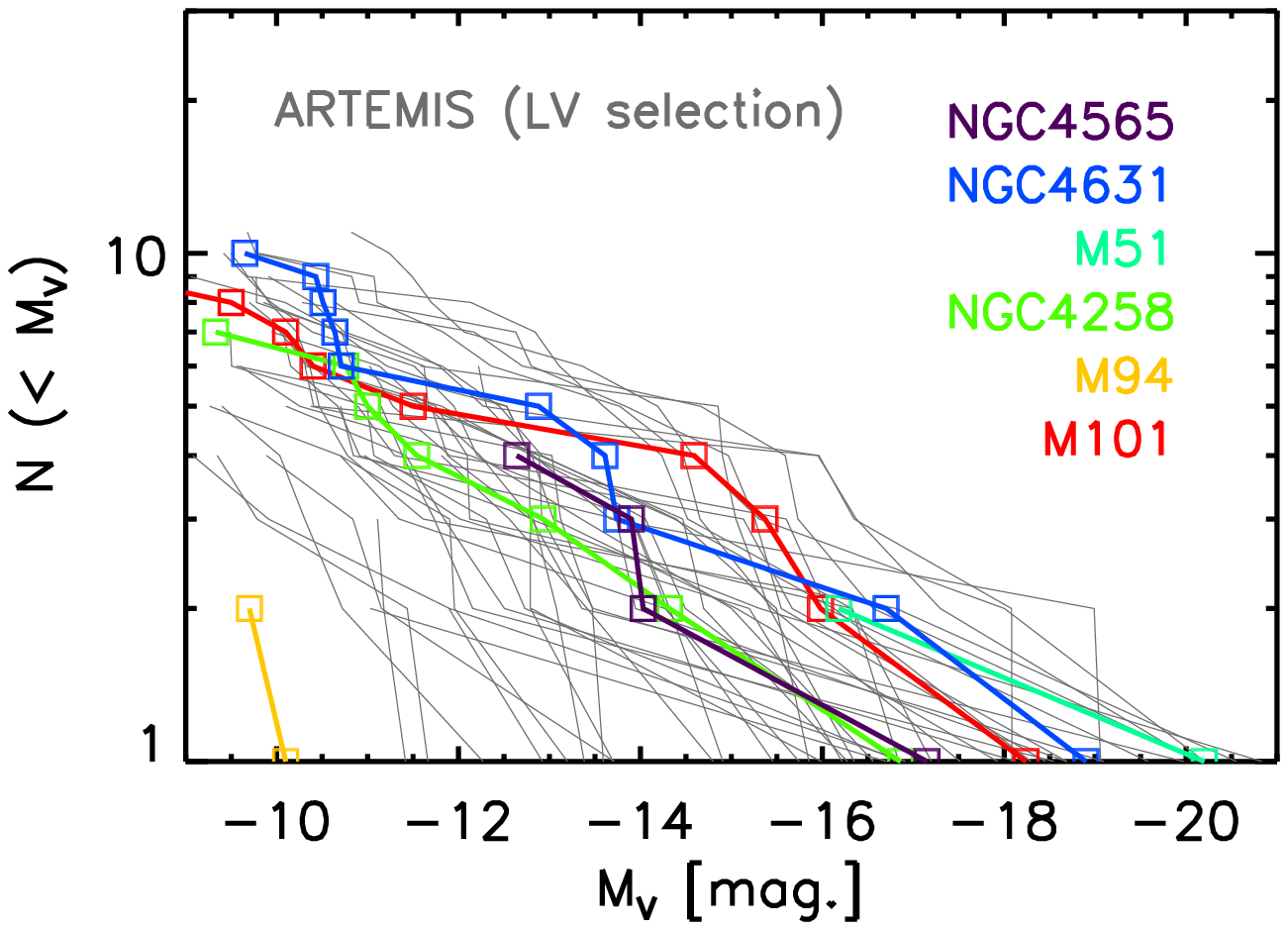} &
  \includegraphics[width=\columnwidth]{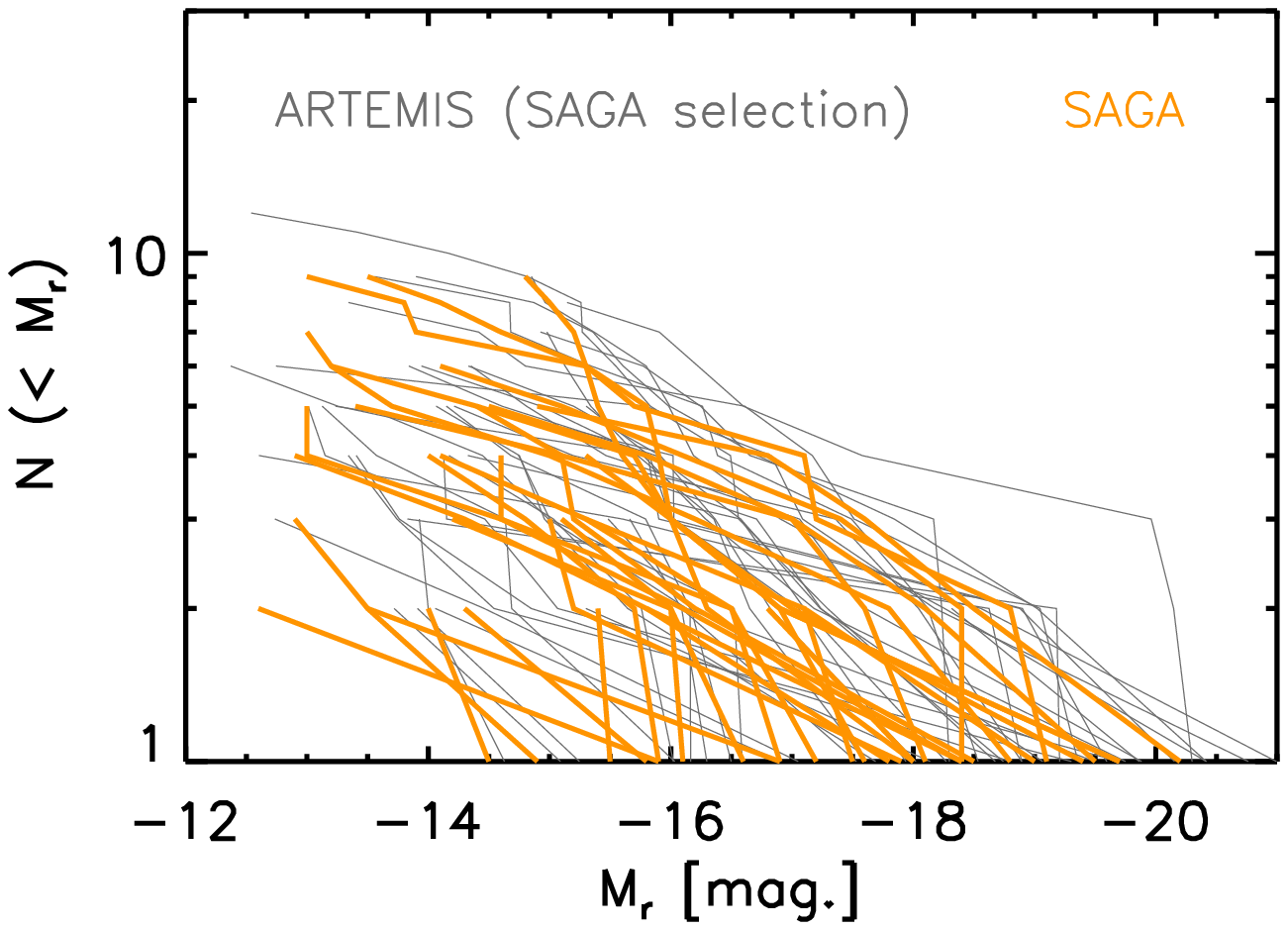}\\
  \includegraphics[width=\columnwidth]{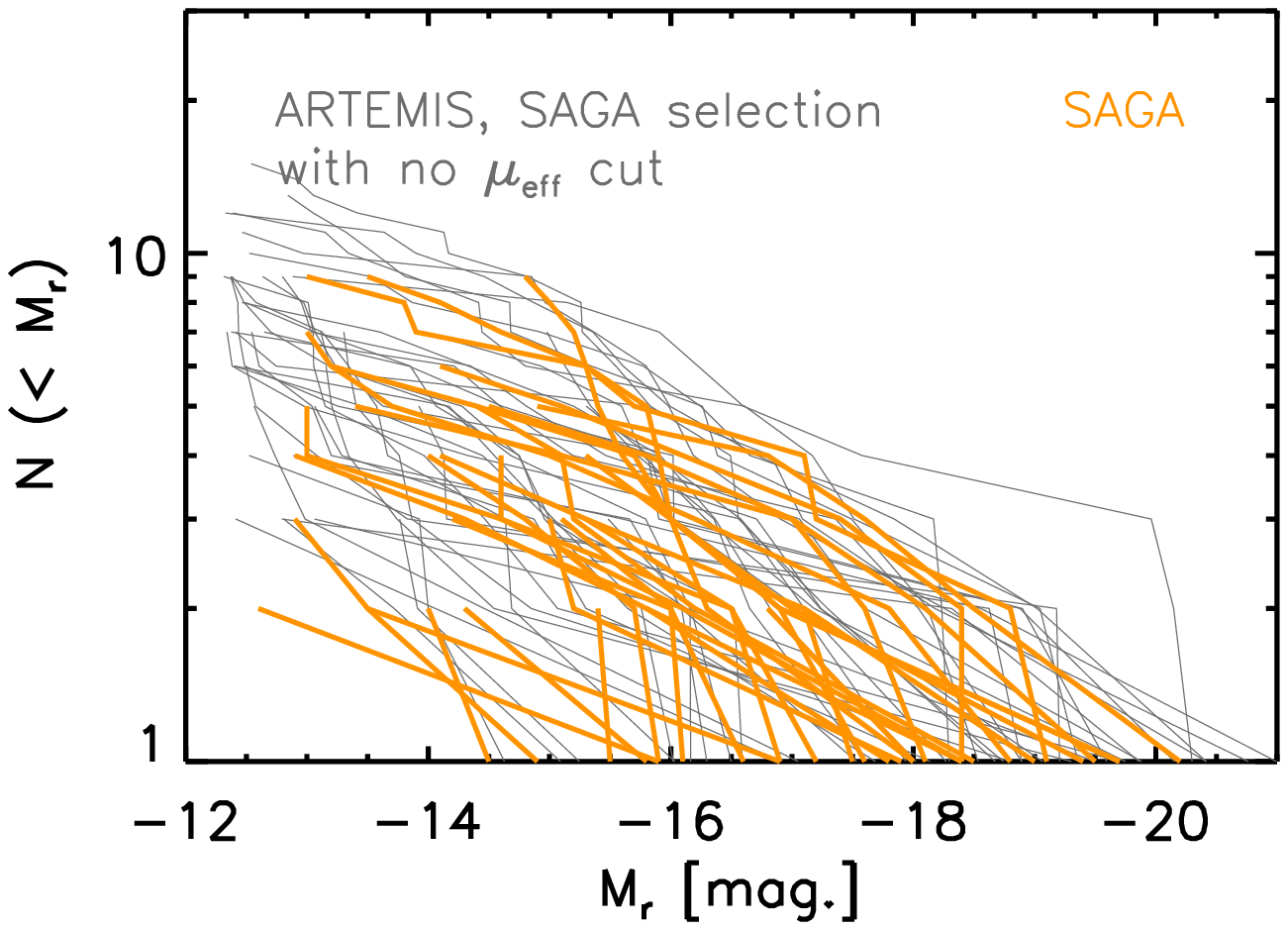} &
  \includegraphics[width=\columnwidth]{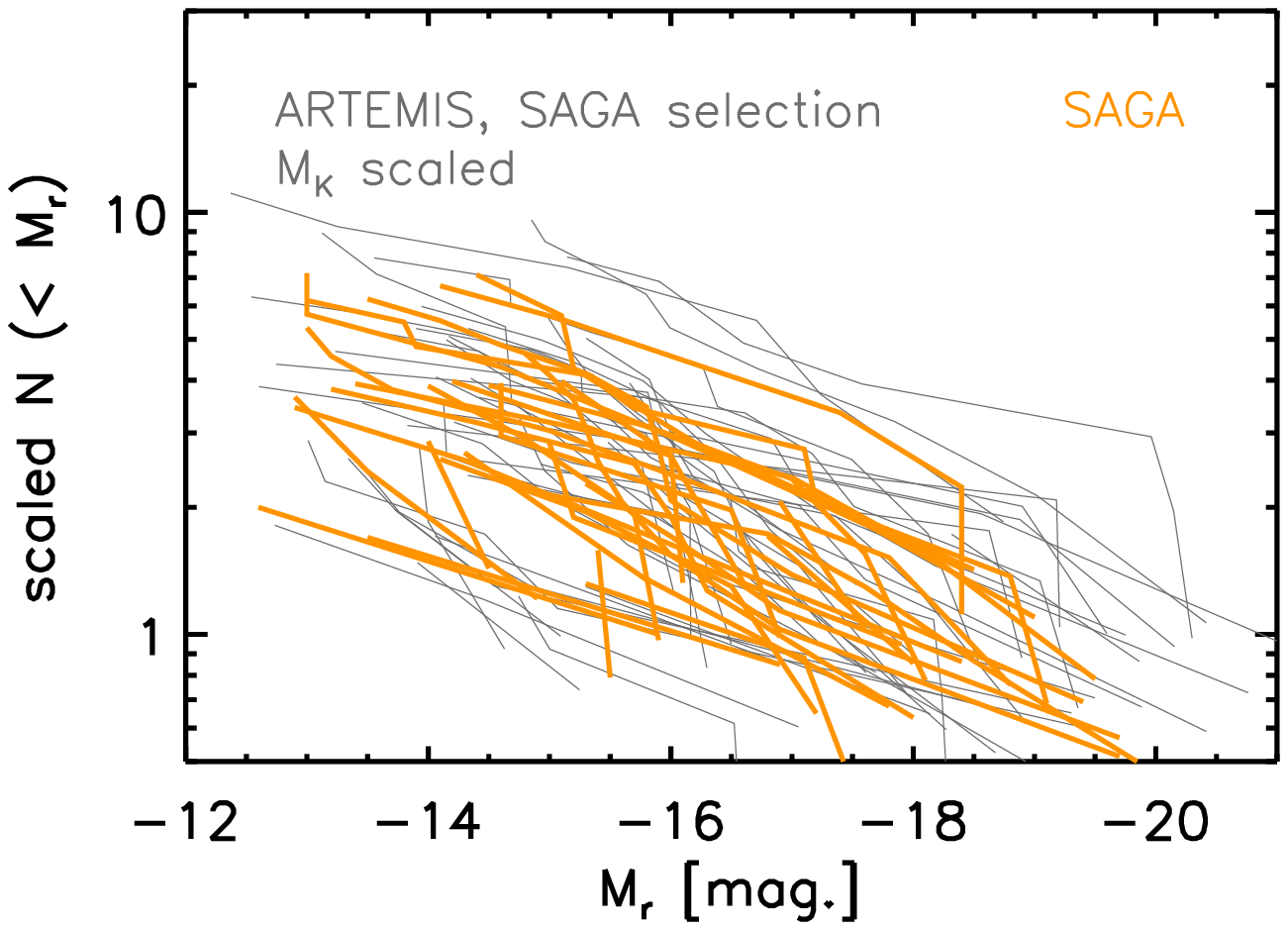}
  \end{tabular}
  \caption{{\it Top left:} $V$-band luminosity functions of all Milky Way-mass galaxies in ARTEMIS (grey lines) in comparison with those of the Milky Way, M31, and of several Milky Way analogs in the Local Volume. {\it Top right:} $r$-band luminosity functions of Milky Way-mass galaxies in ARTEMIS (grey) compared with those of the Milky Way analogues in SAGA \citep{mao2021}. 
  {\it Bottom left:}  Same as top right panel except that no surface brightness cut is applied to the simulations when comparing to the SAGA sample.  {\it Bottom right:}  Same as top right panel except the luminosity functions have been scaled in amplitude to remove the satellite abundance dependence on host K-band magnitude (see bottom right panel of Fig.~\ref{fig:Nsat_scaling} and eqn.~\ref{eq:nsat_mk_saga_sim}), in order to account for small differences in the host magnitude distributions of ARTEMIS and SAGA.}
  \label{fig:lfunc_sat}
\end{figure*}

\subsubsection{Comparisons with Milky Way and M31}

We examine first the simulated satellite cumulative luminosity functions within ${\rm R}_{200}$, separating them by ${\rm M}_{200}$. The top panel of Fig.~\ref{fig:lfunc_MW_M31} shows the $V$-band luminosity functions of all $45$ ARTEMIS galaxies, with dashed and full lines for host masses above and below ${\rm M}_{200} = 10^{12} \, {\rm M}_{\odot}$, respectively. This mass threshold corresponds to ${\rm R}_{200} \approx 200$~kpc, which is close to the median ${\rm R}_{200}$ of our hosts of $\approx 207$~kpc (the full range is between $180-317$~kpc, as shown in Table~\ref{tab:table1}). 

By selecting only subhaloes within ${\rm R}_{200}$ we can compare the number of satellites within a meaningful physical scale in each system. We find, as expected, that more massive host galaxies have more satellites above a fixed brightness. For example, hosts less massive than $10^{12} \, {\rm M}_{\odot}$ in our sample typically contain $5-10$ satellites brighter than $M_V = -8$, whereas more massive hosts typically contain $10-30$ such satellites. This implies that, given the uncertainties in the halo mass of the Milky Way and particularly M31, the simulations predict up to a factor of $\approx 2 - 3$ scatter in the abundance of classical satellites.  A significant scatter has also been found in previous zoom simulations (see, for example, \citealt{sawala2016}), although with a larger sample of Milky Way-mass haloes we can now estimate this scatter more robustly.

Given the uncertainty in the total masses and virial radii of observed systems, it is often customary in observational analyses to include all dwarf galaxies out to some fixed distance, for example within $300$~kpc (e.g., \citealt{mcconnachie2012}). In the middle panel of Fig.~\ref{fig:lfunc_MW_M31} we compare the luminosity functions from the simulations with those of the Milky Way and M31 selecting, for both observations and simulations, all dwarf galaxies within $300$~kpc and with $M_V < -8$. For the observations, we use V-band magnitudes from \citet{mcconnachie2012} and 3D distances calculated by \citet{yniguez2014}. To these, we add measurements of several dwarf galaxies discovered more recently: for M31 we include CasII, CasIII and Lac I \citep{conn2012,martin2013} with updated distances from \citet{weisz2019}, while for Milky Way we include Crater 2 \citep{torrealba2016} and Antlia 2 \citep{torrealba2019}. Note that we include the Sagittarius dwarf and And XIX, even though these are dwarfs in the process of disruption and their counterparts in the simulations may not be easily identified by subhalo finding algorithms (however, our conclusions are not affected by this choice). 

The large scatter in the simulated luminosity functions comfortably brackets the observed luminosity functions for both the Milky Way and M31. This is encouraging, as no aspect of the simulations were calibrated to obtain this result.  However, it still leaves open the question of the precise masses of M31 and the Milky Way. Given its lower (relative) abundance of satellites, the Milky Way appears to require a lower total mass than M31, at least judging on the observed luminosity functions (down to $M_V =-8$, M31 has a factor of $\approx 2$ more satellites than the Milky Way). To demonstrate this more clearly, we determine which simulated systems provide a good match to the luminosity functions of these two galaxies. By experimenting with different sub-sets of our simulations, we find two such sub-sets, one including all hosts with masses ${\rm M}_{200} < 10^{12} \, {\rm M}_{\odot}$ which matches the Milky Way's luminosity function, and another with host masses ${\rm M}_{200} > 1.2 \times 10^{12} \, {\rm M}_{\odot}$, which matches the M31's.  The dashed and dotted lines in the bottom panel of Fig.~\ref{fig:lfunc_MW_M31} show the median luminosity functions of these two sub-sets and the error bars represent the scatter (standard deviation) for those selections.  The subset matching the Milky Way has a median host mass of ${\rm M}_{200} =7.9 \times 10^{11} \, {\rm M}_{\odot}$. In contrast, the sub-set matching the M31 has a median host mass of $1.6 \times 10^{12} \, {\rm M}_{\odot}$. Therefore our results indicate that M31 has a higher mass than the Milky Way (by a factor of $\approx 2$).  We caution, however, that at fixed halo mass there is significant scatter in the luminosity functions and, therefore, precise measurements of the total masses would be difficult to achieve based solely on the abundance of relatively bright satellites.

\subsubsection{Comparisons with Milky Way analogues outside the Local Group}

As discussed in the introduction, a major recent advance in the observations is the ability to identify samples of satellites of Milky Way-mass systems beyond the Local Group. Below we compare the simulated luminosity functions with those of several Milky Way analogues in the Local Volume and out to $\sim 40$~Mpc.

In the top left panel of Fig.~\ref{fig:lfunc_sat} we compare the simulated $V$-band cumulative luminosity functions with those of the Milky Way analogues from the Local Volume, using data from \citet{carlsten2021}, including from these only the distance-confirmed satellites. We apply the `LV selection' criteria discussed in Secton \ref{sec:obs} to the ARTEMIS simulations.
The Local Volume systems are comfortably bracketed by the simulations and the scatter in the observations is similar in magnitude to that for the simulated systems. We note, however, the difference in sample size and the uncertainties in the halo masses of the observed systems. The observations themselves show significant scatter, as noticed previously (see, for example,  \citealt{bennet2019,bennet2020}). Our results indicate that the observed variations in the number of classical satellites can be accommodated within the predictions of the $\Lambda$CDM model, assuming that the observations cover a similar mass range of Milky Way analogues as our simulations. For example, our simulations contain systems as abundant as M31, with more than a dozen satellites within $150$~kpc, but also very sparse systems, like M94, which has only two classical satellites within this radius. This suggests that there is no need to resort to additional scatter in the luminosity functions, such as the stochastic scatter to the stellar mass -- halo mass relation proposed by \citet{smercina2018}, although our results do not rule this possibility out. 

\begin{figure}
  \begin{tabular}{cc}
  \includegraphics[width=\columnwidth]{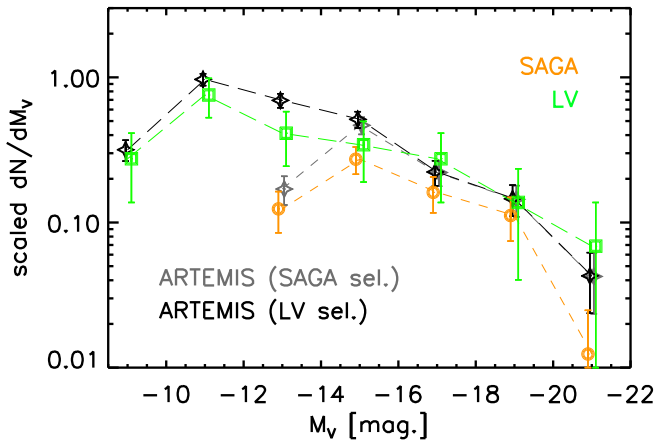} 
  \end{tabular}
  \caption{The mean differential $V$-band luminosity functions of satellites in ARTEMIS (grey), SAGA (orange), and Local Volume (green). For SAGA we have converted the published r-band magnitudes to the V-band assuming a colour $r-V = -0.17$ (the median value from the LV sample). The luminosity functions have been scaled in amplitude to remove the satellite abundance dependence on host K-band magnitude (see bottom right panel of Fig.~\ref{fig:Nsat_scaling} and eqn.~\ref{eq:nsat_mk_saga_sim}), in order to account for small differences in the mean host magnitudes of ARTEMIS, SAGA, and the Local Volume.}
  \label{fig:lfunc_diff}
\end{figure}

\begin{figure*}
\includegraphics[width=6.9in]{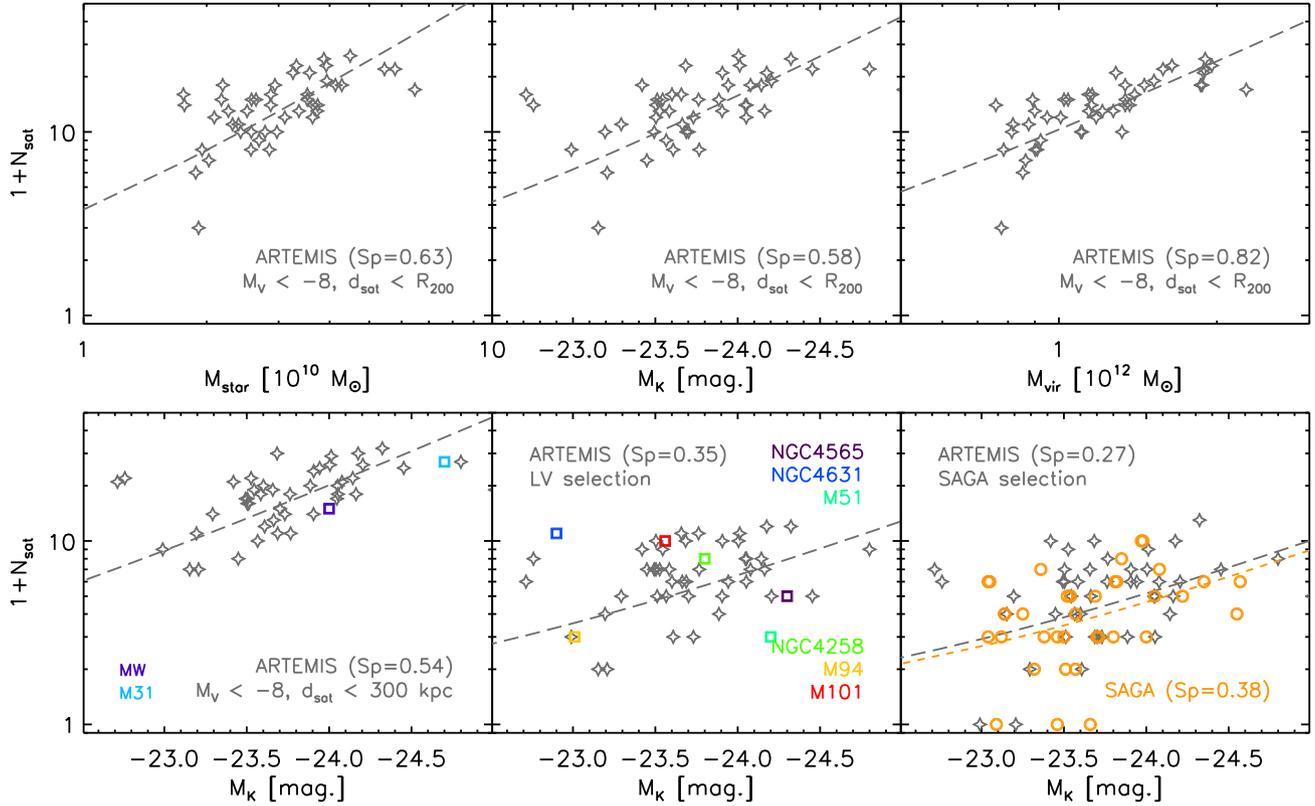}
\caption{Relations between the number of satellites and various properties of their hosts. {\it Top row:} satellites are within the magnitude limit $M_V < -8$ and presently within ${\rm R}_{200}$ from the center of their host. Relations are as a function of: the total host stellar mass (left), versus $M_K$ of the host (middle) and versus virial mass of the host (right). {\it Bottom row:} Relations between ${\rm N}_{\rm sat}$ and host $M_K$ using specific observational constraints. Simulations are shown with empty stars, while coloured squares represent data for the Milky Way (dark blue), M31 (cyan) and other Milky Way analogues (NGC 4565, NGC 4631, M51 and NGC 4258). SAGA  measurements are shown with orange circles.  In all panels the dashed black line represents the best-fit power law relation to ARTEMIS simulations and the dashed orange line in the bottom right panel represents the best-fit power law for the SAGA observations.}
\label{fig:Nsat_scaling}
\end{figure*}

Nevertheless, M94 in the Local Volume survey remains an intriguing system. Among our simulated systems, only a couple of systems, which have halo masses below $10^{12} \, {\rm M}_{\odot}$, are similarly sparse. It is unclear, however, whether M94 has such a low halo mass. \citet{karachentsev2014} calculate a dynamical mass of $\approx 2.6\pm 0.9 \times 10^{12} \, {\rm M}_{\odot}$ when using a standard virial theorem-based estimator.  However, they also explored an alternative `projected mass' estimator from \citet{heisler1985} which is meant to be more robust than standard virial-based techniques.  This approach yielded a total halo mass of only $\approx 2.2 \times 10^{11} \, {\rm M}_{\odot}$ (i.e., an order of magnitude lower than the virial-based estimate).  Such a large discrepancy\footnote{We note that only two of the kinematic tracers used to estimate the masses in \citet{karachentsev2014} were within 300 kpc of M94, with several out to distances of 500 to 700 kpc (i.e., much larger than the likely virial radius), which might help to explain the large difference in the two mass estimates.} in the total mass estimates may reflect the fact that M94 is the central galaxy in a very extended group (the M94 Group).  If the second mass estimate is closer to the truth, the sparsity of M94 can be understood from the strong scaling between satellite abundance and halo mass (see top right panel of Fig.~\ref{fig:Nsat_scaling}), although this would require M94 having a comparatively large host K-band luminosity for its total halo mass.

The top right panel of Fig~\ref{fig:lfunc_sat} compares the $r$-band cumulative luminosity functions of simulated galaxies with those of Milky Way analogues in SAGA \citep{mao2021}, applying the `SAGA selection' criteria in Section \ref{sec:obs} to the ARTEMIS simulations. In agreement with the previous comparison, the scatter in both simulations and observations is relatively large, although this is particularly expected at the bright end of the luminosity function simply as a result of counting (Poisson) errors. Overall, all of the SAGA galaxies in the right panel of Fig~\ref{fig:lfunc_sat} are bracketed by the simulations.

It is noteworthy that individual SAGA and ARTEMIS (with SAGA selection) luminosity functions often do not extend down to the magnitude limit of $M_r = -12.3$.  This is a consequence of the surface brightness limit.  To demonstrate this, in the bottom left panel of Fig~\ref{fig:lfunc_sat} we remove the surface brightness limit from the ARTEMIS satellite selection criteria, which preferentially allows more low-luminosity systems to be included and many of the ARTEMIS luminosity functions now extend right down to the magnitude limit.  Note that the agreement with SAGA for luminosities brighter than $M_v \approx -14$ is unaltered as a result of dropping the surface brightness limit.  Thus, an alternative approach to implementing both a magnitude and a surface brightness cut in the simulations is to limit the comparison to relatively bright satellites.  Furthermore, we note that the impact of the surface brightness cut on the comparison to the Local Volume sample (top left panel) is less dramatic relative to the SAGA comparison, in that many of the observed and simulated systems (with LV selection) extend to the magnitude limit of $M_V = -9$ even in the presence of a surface brightness limit.

As discussed in Section \ref{sec:hosts}, the host K-band magnitude distributions of ARTEMIS, SAGA, and LV are very similar, but they are not identical.  It is therefore of interest to examine whether differences in the host magnitude distributions, small though they may be, bias the comparisons to the observations.  With this in mind, in the bottom right panel of Fig~\ref{fig:lfunc_sat} we show the comparison to SAGA survey but now with the dependence on host K-band magnitude scaled out.  To achieve this, we apply a correction factor based on eqn.~\ref{eq:nsat_mk_saga_sim} (discussed below) to remove the dependence of satellite abundance on host K-band magnitude.  Specifically, for each individual system in ARTEMIS and SAGA we scale (divide) the abundance by a factor $(10^{[M_K + 23.5]})^{-0.34}$, which is the ratio of the abundance expected for a given host K-band magnitude from the best-fit power law (to the K-band magnitude--satellite abundance relation in Fig.~\ref{fig:Nsat_scaling}) to that at the pivot point.  Thus, systems with large (small) K-band luminosities relative to the pivot point ($M_K = -23$) have their abundances scaled down (up).   Note that, as the correction is a scale factor applied to the whole luminosity function, it is implicitly assumed that the \textit{shape} of the luminosity function does not depend on host K-band magnitude over the range of magnitudes explored here.  Overall, the level of agreement between ARTEMIS and SAGA is virtually unchanged compared with the top right panel of Fig~\ref{fig:lfunc_sat}, owing to the similarity of their host K-band magnitude distributions and the relatively weak dependence of satellite abundance on K-band host magnitude (and its large intrinsic scatter) when observational selection criteria are applied.

It is interesting to note that both \citet{mao2021} and \citet{carlsten2021} reported evidence for a slight excess of bright satellites ($\la -18$ mag.) with respect to the simulations they compared to, which were based on dark matter-only simulations populated using abundance matching techniques.  We examine whether this is the case here by comparing the mean differential $V$-band luminosity functions of ARTEMIS, SAGA, and the Local Volume in Fig.~\ref{fig:lfunc_diff}.  We convert the $r$-band luminosities of SAGA into the $V$ band using $r-V = -0.17$ and we scale the luminosity functions in amplitude to account for the small differences in mean host $K$-band magnitude.  Two different luminosity functions are computed for ARTEMIS, corresponding to the different selection criteria for SAGA and LV.  Note also that for this comparison we select only satellites within 150 kpc of the host, to ensure a fair comparison between the three data sets.  We find generally good agreement (within the $1$-sigma Poisson uncertainties) with both SAGA and LV.  There is no obvious sign of an excess of bright satellites in the observational data sets with respect to ARTEMIS.  Fig.~\ref{fig:lfunc_diff} also nicely shows the impact of the different selection criteria for magnitudes fainter than $M_V \approx -14$.  As noted by \citet{carlsten2021}, there is a difference between SAGA and the Local Volume at magnitudes fainter than this limit, which we attribute to a surface brightness limit in SAGA.

In terms of the bright-end of the luminosity function, it is presently unclear what the origin of the difference is between the predictions of ARTEMIS and that of abundance matching examined in \citet{mao2021} and \citet{carlsten2021}.  \citet{carlsten2021} discuss two possible explanations for the difference between the abundance matching-based predictions and their observations, namely that: i) the intrinsic scatter in stellar mass at fixed halo mass could be larger than assumed in the fiducial abundance matching analyses; and/or ii) the local galaxy stellar mass function (from GAMA) used in the abundance matching analyses is slightly incomplete at stellar masses of $\sim10^9$ M$_\odot$ (e.g., \citealt{Sedgwick2019}).  Either of these options could also explain the difference between the predictions of ARTEMIS and abundance matching.  Clearly a more careful comparison between the predictions is required in order to ascertain the origin of the difference, which we leave for future work.

\begin{figure*}
\begin{tabular}{cc}
\includegraphics[width=\columnwidth]{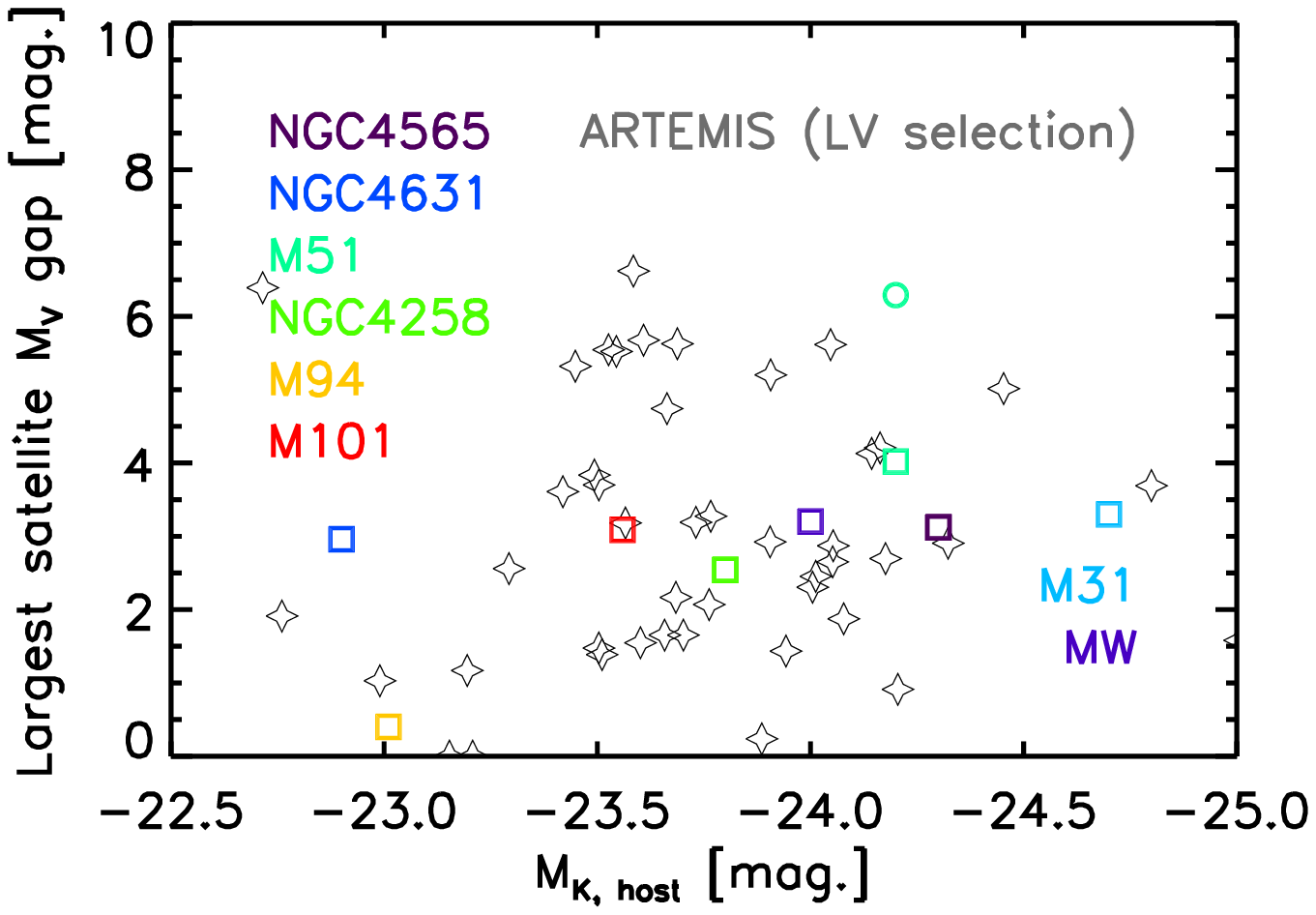} &
\includegraphics[width=\columnwidth]{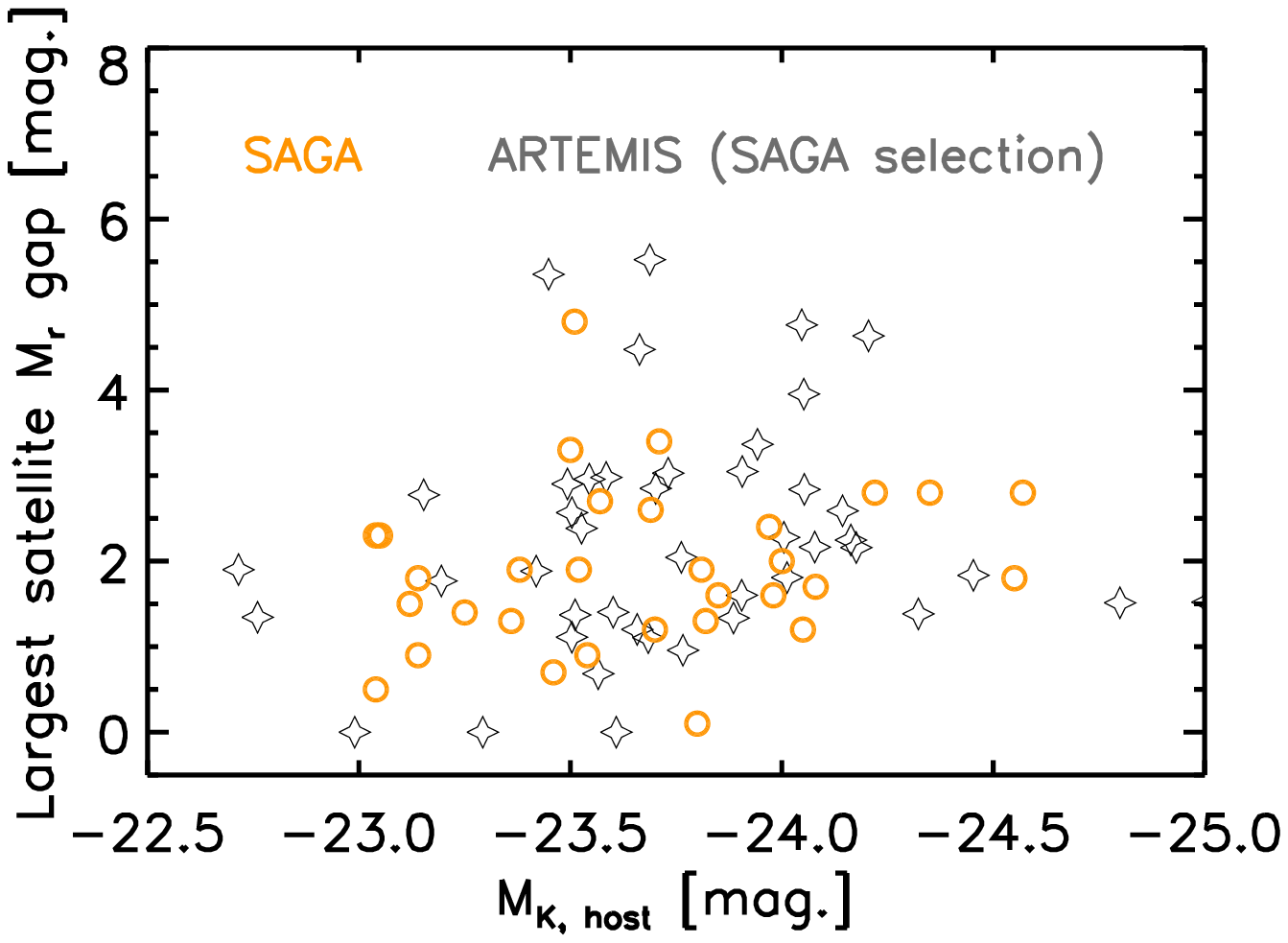}
\end{tabular}
\caption{The largest magnitude gaps in the simulated luminosity functions (with star symbols) compared with those in the Local Volume data ({\it left panel}) and in SAGA ({\it right panel}), using the specifics of each survey. The observational data are from \citet{carlsten2021} and \citet{mao2021}, respectively. We also include measurements from the Milky Way and M31, from  sources mentioned earlier. The largest magnitude gaps are plotted versus the magnitudes of the host, in the $V$-band and $K$-band, respectively. The range in $M_V$ and $M_r$ gaps in the simulations is similar to that in the observations.}
\label{fig:sat_gap}
\end{figure*}

Moving on, the slopes of the luminosity functions, typically measured via the successive magnitude gaps, provide additional information about satellite populations. Prior measurements of the largest magnitude gaps in the luminosity functions have suggested another potential problem for the $\Lambda$CDM model. Specifically, a few of the largest magnitude gaps measured in the SAGA `stage 1' sample \citep{geha2017} appear to be larger than the $2\sigma$ scatter around the mean predicted by theoretical models.  On the other hand, it has been argued that, statistically, such large magnitude gaps are expected in a $\Lambda$CDM cosmology \citep{ostriker2019}.
 
In Fig.~\ref{fig:sat_gap} we investigate the largest magnitude gaps in the simulations and compare with the observations, namely the Local Volume, SAGA, and the Milky Way and M31. Here the largest magnitude gap is defined as the gap between adjacent $M_V$ (or $M_r$) values of dwarf galaxies (hence, excluding the host). The values for simulations are shown with star symbols and various values measured in observations are shown with coloured squares. Overall, the simulated values compare well with both sets of observations. The medians of the largest magnitude gaps in the simulations are $2.69$ for $V$-band and $1.83$ for $r$-band (note though that the two sets of observations have different magnitude limits and spatial coverage of dwarf galaxies). The median value of the largest $M_r$ gaps in our simulations is very good agreement with the median value of the models used by \citet{geha2017}, of $\approx 1.8$~mag. However, the scatter around the median largest $M_r$ gap in our simulations is larger than the $2 \sigma$ scatter in the models used by \citet{geha2017}.  While understanding the differences between the scatter in these models and in simulations is certainly useful, we note that the discrepancy between the former and the observations may be alleviated with the latest (SAGA) data. The larger SAGA  sample (compared with the stage 1 SAGA sample presented in \citealt{geha2017}) contains several systems with large magnitude gaps, including one with a gap $>4.5$ mag, which was the upper $95\%$ limit in the models used in \citet{geha2017}.  

As for the largest $V$-band magnitude gaps, some of the values in the simulations can exceed $6$~mag, with no counterparts yet in the Local Volume sample with confirmed satellites only (coloured squares on the left panel of Fig.~\ref{fig:sat_gap}). However, by including other `possible' satellites from \citet{carlsten2021}, it appears that M51 may possess such a large gap (green circle in the same panel).

Finally, we note that the comparisons in Fig.~\ref{fig:sat_gap} applied the `LV selection' and `SAGA selection' criteria to the simulations.  Virtually identical results are obtained if only a magnitude cut is applied (and no surface brightness cut), as the largest magnitude gaps are almost always determined at the bright end of the luminosity functions where the surface brightness limit is unimportant.

\subsection{Abundance scaling relations}
\label{sec:abundance}

Having explored the luminosity functions of satellite around Milky Way-mass galaxies, we now examine how the abundance of satellites (i.e., their total number above some luminosity limit) correlates with the properties of the central host galaxy and with that of the overall host halo.

The top three panels of Fig.~\ref{fig:Nsat_scaling} show the relations between the abundance of satellites, defined here as ${\rm N}_{\rm sat} (M_V <-8)$, and the total stellar mass (computed within a fixed aperture of $30$~kpc), the $K$-band magnitude, which is expected to correlate well with the stellar mass, and the virial mass of the host. Only subhaloes within ${\rm R}_{200}$ of the center of the host are included in the top row. As expected, we find a strong correlation with the host ${\rm M}_{\rm vir}$.  The computed Spearman rank correlation coefficient is $0.82$. (A similar factor is found if we include dwarf galaxies within $300$~kpc of their host.)  
The correlation with host stellar mass (${\rm M}_{\rm star}$) is also significant, with a Spearman coefficient of $0.62$.  Physically, we expect the abundance of haloes to be more strongly tied to the overall halo rather than with the central galaxy and the scatter in the stellar mass -- halo mass relation is responsible for the somewhat weaker correlation between the satellite abundance and the host stellar mass.  The correlation ${\rm N}_{\rm sat} - M_K$ is similar in strength with that of ${\rm N}_{\rm sat} - {\rm M}_{\rm star}$ (its Spearman coefficient is $0.58$), reflecting the known sensitivity of this magnitude band to the total stellar mass.  

The dashed black lines in top row of Fig.~\ref{fig:Nsat_scaling} represent the best-fit power law distributions to the abundance scaling relatons for satellites within ${\rm R}_{200}$ and with $M_V \le -8$.  Specifically, we find:

\begin{equation}
\label{eq:nsat_mstar}
N_{\rm sat}(<{\rm R}_{200}) = (11.95 \pm 0.54) \biggl(\frac{M_{\rm star}}{3\times10^{10}~{\rm M}_\odot}\biggr)^{1.33 \pm 0.18} \ \ \ ,
\end{equation}

\begin{equation}
\label{eq:nsat_mk}
N_{\rm sat}(<{\rm R}_{200}) = (8.81 \pm 0.59) \biggl(10^{[M_K + 23.5]}\biggr)^{-0.45\pm0.06} \ \ \ ,
\end{equation}

\begin{equation}
\label{eq:nsat_mvir}
N_{\rm sat}(<{\rm R}_{200}) = (9.31 \pm 0.54) \biggl(\frac{M_{\rm vir}}{10^{12}~{\rm M}_\odot}\biggr)^{1.32 \pm 0.15} \ \ \ , 
\end{equation}

\noindent for the three scaling relations, where the best-fit parameters were determined using a simple $\chi^2$ minimisation scheme and incorporating Poisson errors on the simulation abundances.

In the bottom three panels of Fig.~\ref{fig:Nsat_scaling} we apply various observationally-motivated choices and limitations. In the left panel we plot again ${\rm N}_{\rm sat} - M_K$, but now counting all dwarf galaxies out to $300$~kpc and brighter than $M_V=-8$. The correlation does not change significantly compared with the case above, when only systems within ${\rm R}_{200}$ were considered (the Spearman coefficient is now $0.54$, compared with $0.58$ previously). The best-fit power law to this relation is:
\begin{equation}
\label{eq:nsat_mk_300}
N_{\rm sat}(<300 {\rm kpc}) = (12.27 \pm 0.71) \biggl(10^{[M_K + 23.5]}\biggr)^{-0.39\pm0.05} \ \ \,
\end{equation}

In the middle and right panels we apply the LV and SAGA selection criteria described in Section \ref{sec:obs}.  With these constraints, the simulated and observed ${\rm N}_{\rm sat} - M_K$ relations agree with each other, but the correlations are considerably weaker. The Spearman coefficients in the simulations are $\approx 0.35$, in approximate agreement with the Spearman coefficient of $\approx 0.38$ obtained for the SAGA sample (see also \citealt{geha2017}). 

\begin{figure*}
\begin{tabular}{cc}
     \includegraphics[width=\columnwidth]{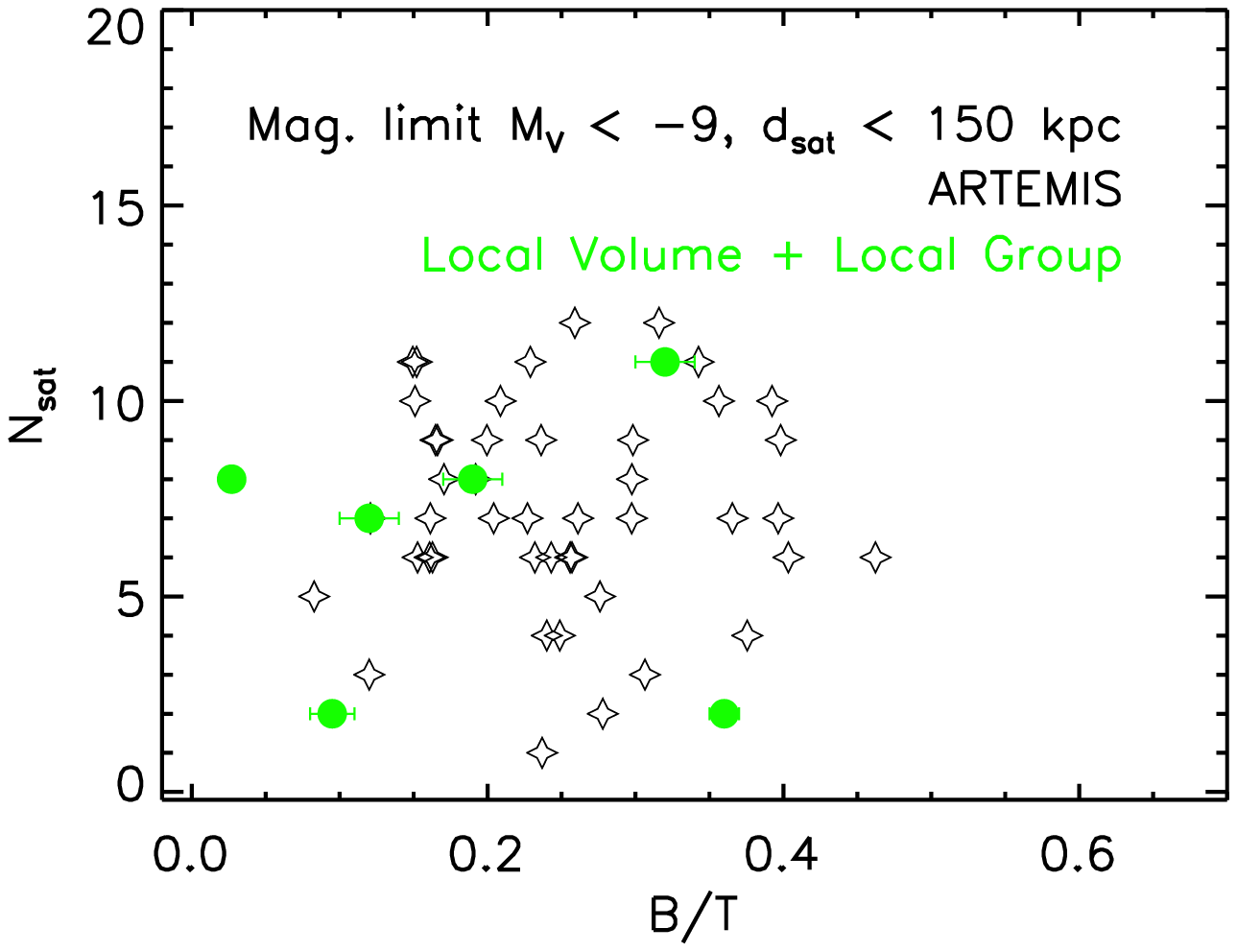} &
     \includegraphics[width=\columnwidth]{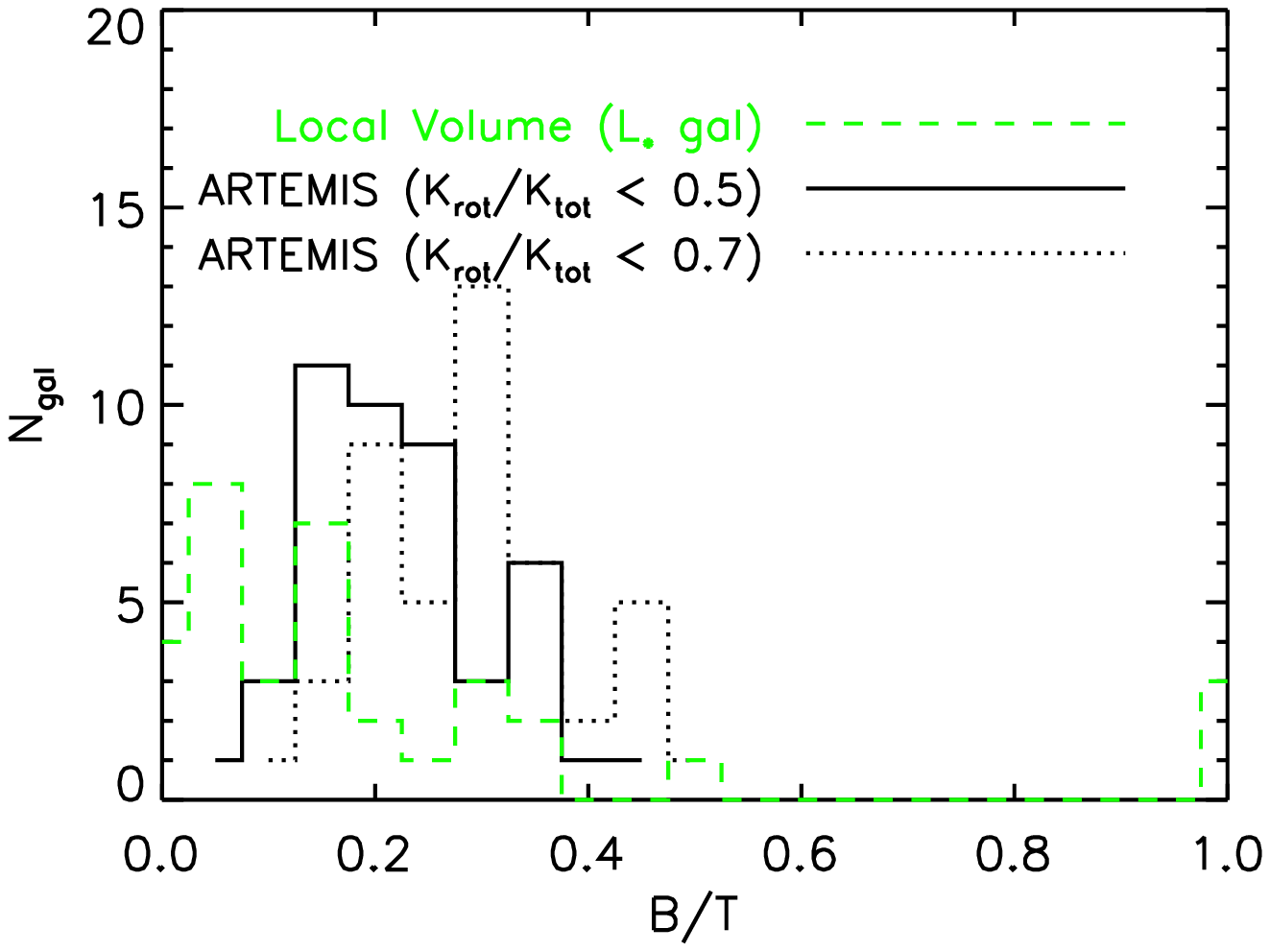} 
     \end{tabular}
         \caption{Left: The number of bright ($M_V <-9$) satellites versus the bulge-to-total ratio, $B/T$, of their host galaxies in the simulations (empty stars) versus similar measurements in the Local Group and Local Volume (filled circles).  For consistency with Local Volume data, we include only satellites with $M_V <-9$ out to $150$~kpc from the center of their host. Observations include the Milky Way, M31, NGC4258, M51, M94 and M101, with $B/T$ values from \citet{kormendy2010} and  \citet{fisher2011}. ${\rm N}_{\rm sat}$ values are from the sources mentioned earlier. {\it Right:} The distribution of $B/T$ values in ARTEMIS, in comparison with $34 \, L_*$ galaxies in the Local Volume and Local Group (from \citealt{peebles2020} and references therein). For ARTEMIS, we use two cuts in ${\rm K}_{\rm rot}/{\rm K}_{\rm tot}$ to select the bulge (see text). Both simulated and observed $B/T$ values are strongly clustered around small values.}
    \label{fig:Nsat_BT}
\end{figure*}

The best-fit power law to the simulations when applying the LV selection criteria is:
\begin{equation}
\label{eq:nsat_mk_lv_sim}
N_{\rm sat,~LV~sel.} = (3.75 \pm 0.37) \biggl(10^{[M_K + 23.5]}\biggr)^{-0.33\pm0.09} \ \ \ ,
\end{equation}

\noindent whereas the best-fit power law when applying the SAGA selection criteria to ARTEMIS is:
\begin{equation}
\label{eq:nsat_mk_saga_sim}
N_{\rm sat,~SAGA~sel.} = (2.84 \pm 0.36) \biggl(10^{[M_K + 23.5]}\biggr)^{-0.34\pm0.12} \ \ \ .
\end{equation}

We also derive the best-fit power law from the SAGA data itself, finding:
\begin{equation}
\label{eq:nsat_mk_saga}
N_{\rm sat,~SAGA} = (2.47 \pm 0.31) \biggl(10^{[M_K + 23.5]}\biggr)^{-0.34\pm0.14} \ \ \ ,
\end{equation}

\noindent where when fitting the SAGA data or the simulations with SAGA selection criteria applied we exclude the small number of hosts with no satellites.  The ARTEMIS simulations reproduce the observed SAGA relation both in slope and amplitude (to within $\approx 1$-sigma).

We highlight that the implementation of a surface brightness cut for the ARTEMIS satellites is important for the comparison to SAGA.  Without a surface brightness cut, the best-fit power law becomes:
\begin{equation}
\label{eq:nsat_mk_saga_sim_mag_only}
N_{\rm sat,~SAGA~mag.~only} = (3.63 \pm 0.40) \biggl(10^{[M_K + 23.5]}\biggr)^{-0.45\pm0.10} \ \ \ .
\end{equation}
\noindent which is an increase of 30\% in abundance at the $M_K = -23.5$ relative to the case without a surface brightness cut, and larger than this for brighter systems (note the steeper trend when the surface brightness cut is neglected).

These results demonstrate that, by limiting the observations to the brightest of dwarf galaxies, the correlations between the number of satellites and host galaxy properties are much less evident.  It also suggests that observations with significant limitations (either in magnitude or in physical extent) typically have a higher scatter in the abundance of satellites at fixed host properties than the intrinsic scatter predicted by cosmological models that do not impose such restrictions.

In the left panel of Fig.~\ref{fig:Nsat_BT} we investigate the relation between the number of satellites and the bulge-to-total ratios of the host galaxies. Such a relation was suggested by \citet{javanmardi2020}, who found a very strong correlation between these two parameters (with a Spearman coefficient of $0.9$), although on a very small sample of galaxies.  For our comparison, we include four Milky Way-mass analogues from the Local Volume sample of \citet{carlsten2021} with complete samples of bright ($M_V<-9$) satellites  within $150$~kpc of their hosts, and with measured $B/T$ values \citep{kormendy2010,fisher2011}, to which we add similar data for the Milky Way and M31. For the simulations, we compute $B/T$ ratios using a kinematical decomposition of the bulge/halo and the disc. Specifically, we assign to the bulge only the star particles within the inner $5$~kpc and with no significant rotation, i.e. with a fraction of energy in rotation of less than $50\%$ of the total energy, i.e. ${\rm K}_{\rm rot}/{\rm K}_{\rm tot} <0.5$. We note however that the $B/T$ ratios computed by kinematic decomposition may differ from the values determined in observations, typically using decomposition of light profiles. 

We do not find any significant correlation between ${\rm N}_{\rm sat}$ and $B/T$, either in  the simulations or the observations (the Spearman coefficients are essentially $0$). We checked that this result is not sensitive to the exact luminosity or spatial cuts; for example, if we select simulated satellites with $M_V<-8$ and within a distance ${\rm R}_{200}$ of their hosts, we still find a low Spearman coefficient (of $0.2$). We note that \citet{javanmardi2019} also investigated such correlations using the Illustris simulations and have not found a correlation between the abundance of satellites and the host galaxy's $B/T$. Rather than this being a problem with the $\Lambda$CDM model, as suggested by \citet{javanmardi2020}, our results suggest that the observed ${\rm N}_{\rm sat} - B/T$ relation can be easily skewed by including a few systems with extreme values (e.g. systems like M33 with $B/T\approx 0$ or Cen A with $B/T \approx 1$, neither of which are of Milky Way-mass). 

The right panel of Fig.~\ref{fig:Nsat_BT} shows the $B/T$ distribution for the simulated galaxies versus the distribution in the Local Volume. For the simulations, we show two cuts in the kinematic selection of bulge stars: one with ${\rm K}_{\rm rot}/{\rm K}_{\rm tot} <0.5$  for stars in the inner $5$~kpc (this is the same cut as used in the left panel of Fig.~\ref{fig:Nsat_BT}), the other with ${\rm K}_{\rm rot}/{\rm K}_{\rm tot} <0.7$, respectively. Both choices are somewhat arbitrary and will not correspond exactly with the bulge/disc decomposition based on light profiles more commonly employed in observations.  Including stars with energies dominated by rotation is likely to contaminate the bulge with disc stars (as shown in fig.~B1 of \citealt{font2020}, stars with ${\rm K}_{\rm rot}/{\rm K}_{\rm tot}>0.8$ are located in star-forming regions). For the observational sample, we include all $L \sim L_*$ galaxies in the \citet{tully2009} catalogue\footnote{See the `Local Universe' catalogue at the Extragalactic Distance Database, http://edd.ifa.hawaii.edu.} with measured $B/T$ values, plus the Milky Way and M31 ($34$ galaxies in total). We use the $B/T$ values from Table~1 of \citet{peebles2020}, which are measurements from \citet{kormendy2010} and \citet{fisher2011}. The simulated and observed distributions are in reasonably good agreement (note that the samples are also of similar size). All distributions have a narrow range in $B/T$ which suggests a correlation with ${\rm N}_{\rm sat}$ - which it has been found to display a large scatter - to be unlikely.

Our simulations obtain a low median $B/T$ value, particularly when the bulge selection is limited to stars without significant rotation. This is encouraging, although we caution that the result needs to be confirmed with a bulge/disc decomposition based on light profiles, as in observations. Cosmological simulations have consistently encountered problems in producing disc galaxies with small bulges (see, e.g. \citealt{brooks2016} and references therein). \citet{peebles2020} has recently compared the $B/T$ values in the Local Volume galaxies (included by us above) with those in the Auriga \citep{gargiulo2019} or FIRE-2 \citep{garrison-kimmel2018} simulations, and shown that the problem still exists in these simulations. The $B/T$ values in ARTEMIS are in the same range as in those simulations if we use the ${\rm K}_{\rm rot}/{\rm K}_{\rm tot} <0.7$ cut, but significantly lower if we use the ${\rm K}_{\rm rot}/{\rm K}_{\rm tot} <0.5$ selection. This suggests that the nature of the kinematic threshold is important in the bulge selection in simulations. We plan to investigate this issue in more detail in a future study. 
 
\subsection{Radial distribution of dwarf galaxies around Milky Way analogues}
\label{sec:radial}

We now examine the radial distribution of dwarf galaxies around their hosts.  As in previous sections, we begin by comparing the simulations to the Milky Way and M31 before moving on to the Milky Way analogues beyond the Local Group.

\subsubsection{Comparison with the Milky Way and M31}
\label{sec:radial_MW_M31}

\begin{figure}
\includegraphics[width=\columnwidth]{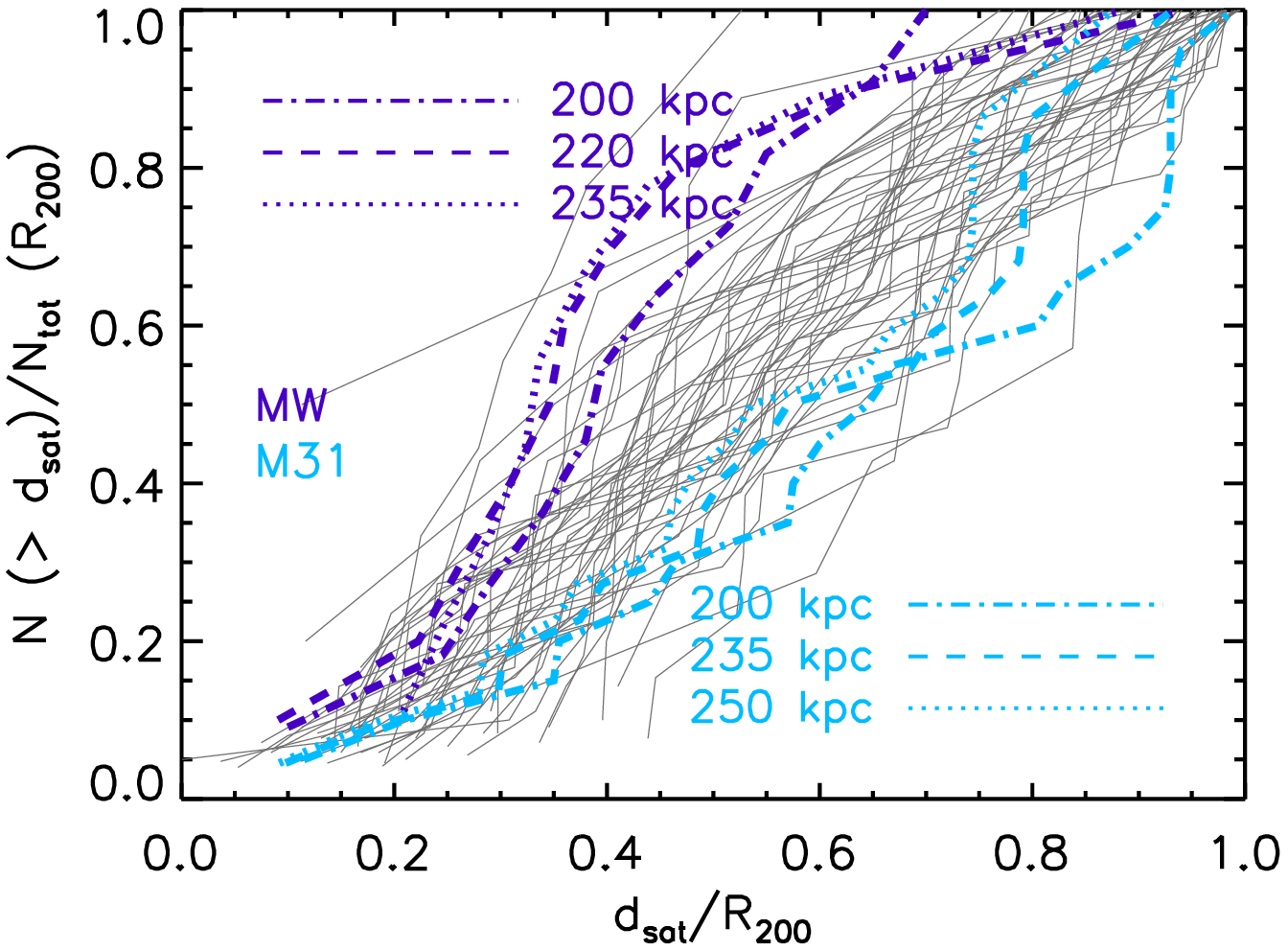}
\includegraphics[width=\columnwidth]{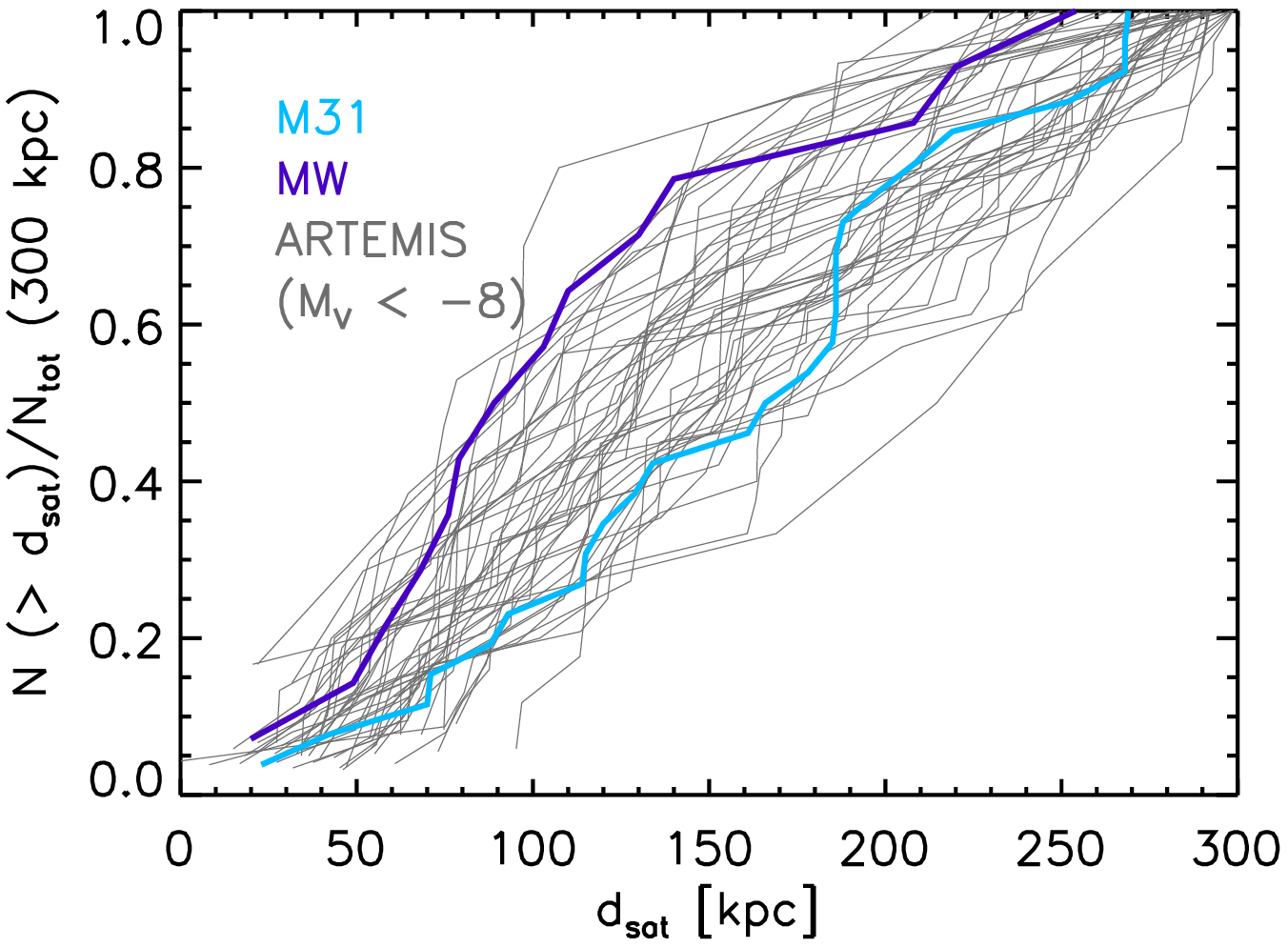}
\caption{{\it Top:} Radial distribution of bright satellites ($M_V <-8)$), in terms of their 3D distance from the center of the host, $d_{\rm sat}$ normalised by $R_{200}$, in the ARTEMIS simulations (grey lines).  For the observations (various purple curves for the Milky Way and cyan curves for M31), we adopt different plausible values for $R_{200}$ of the Milky and M31, based on recent total mass measurements of these systems.  {\it Bottom:} Radial distributions out to a fixed radius of $300$~kpc in ARTEMIS (grey lines), Milky Way (full blue line), and M31 (full cyan line).}
\label{fig:rad_distrib}
\end{figure}

With the ARTEMIS simulations, we can now revisit the claimed tension (discussed in the introduction) between the predictions of $\Lambda$CDM models for radial distributions of satellites of Milky Way-mass galaxies and the measured radial distributions. As already mentioned, the Milky Way appears to be a rare system, with the majority ($80\%$) of its classical satellites (out to $300$~kpc) being clustered in the inner $100$~kpc or so \citep{yniguez2014}.  From a volume point of view, the majority of classical satellites apparently reside within central $10-20\%$ of the volume contained within ${\rm R}_{200}$.

The top panel of Fig.~\ref{fig:rad_distrib} shows the radial distributions of simulated satellites, i.e systems within ${\rm R}_{200}$ with $M_V < - 8$ in ARTEMIS (grey lines) compared with those of Milky Way and M31 (blue and cyan lines, respectively). For observations we use the same dwarf galaxies mentioned in Section \ref{sec:lumfunc}. Since we do not know the precise values for ${\rm R}_{200}$ for the Milky Way and M31, we adopt a range of plausible values given the range of total mass estimates in the recent literature.  Interestingly, we find that both the Milky Way and M31 (the former being significantly more concentrated than the latter) lie within the bounds of the simulated systems, even though they tend to exist on the extremities of the simulated distribution. 

In the bottom panel of Fig.~\ref{fig:rad_distrib} we compare the radial distributions within a fixed distance of $300$~kpc. For both simulations and observations, we include only satellites with $M_V <-8$.  Here both the Milky Way and M31 again fall within the scatter of the simulations.
The inclusion of Crater 2 and Antlia 2 -- which lie at distances of $\approx 110$~kpc \citep{torrealba2016} and $\approx 130$~kpc \citep{torrealba2019}, respectively -- alleviates some of the previously claimed tension between the Milky Way and previous simulations. With these additions, $\approx 60\%$ of Milky Way's classical dwarfs (within $300$~kpc) are now within $100$~kpc. 

It is also possible that a few classical dwarf galaxies could remain undetected observationally.   For example, the observations do not include many systems like Crater 2 and Antlia 2, which have effective surface brightnesses $>30$ mag arcsec$^{-2}$ \citep{torrealba2016,torrealba2019} and were discovered only recently.

\begin{figure}
\includegraphics[width=\columnwidth]{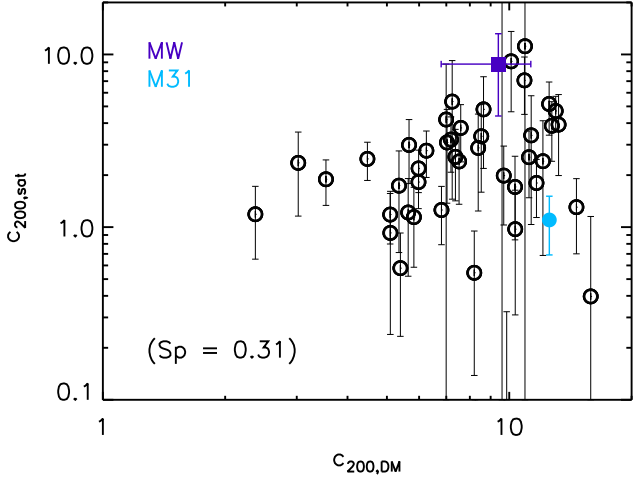}
\caption{The relation between the concentration of the satellite population, $c_{200, {\rm sat}} \equiv {\rm R}_{200}/r_{s, {\rm sat}}$ (where $r_{s, {\rm sat}}$ is the NFW scale radius of the satellite radial distribution, see text) and the concentration of dark matter halo of the simulated host galaxies, $c_{200, {\rm DM}}$.  The correlation is relatively weak (Spearman coefficient of $\approx 0.31$), but there is a tendency for haloes that are highly concentrated (and that formed earlier) to also have more radially concentrated satellites.}
\label{fig:c200}
\end{figure}

Assuming that the number of undetected ultra-diffuse classical dwarfs is not large, how can we explain the difference between the radial distributions of Milky Way and M31 satellites?  Moreover, our simulations indicate a large scatter in the radial distributions of dwarf galaxies, in a relatively narrow host mass range, even when we have normalised the radial coordinate by ${\rm R}_{200}$ and have therefore removed the first order dependence of halo mass.  Is this scatter purely stochastic, or does the scatter depend (at least in part) on other properties of the host? In Fig.~\ref{fig:c200} we examine how the concentration of the satellite system (defined below) depends on the concentration of host dark matter halo, $c_{200, {\rm DM}}$.  We have also examined the dependence of the satellite concentration on the halo formation age, $t_{\rm form}$ (i.e., the lookback time to the formation redshift the halo, which is defined below).  The halo concentration and halo formation time are known to be strongly correlated with each other and provide simple measures of the formation history of halo, with more highly-concentrated haloes tending to have assembled their mass earlier on (e.g., \citealt{navarro1996,wechsler2002}).  For brevity, we do not show the trend with $t_{\rm form}$ in Fig.~\ref{fig:c200}, but we find it to be similar (in terms of the strength of the correlation) to the trend with the dark matter concentration.

The host dark matter concentrations are estimated by first producing spherically-averaged density profiles of the dark matter (excluding the contribution of satellites) and fitting an NFW form to the density profiles over the radial range $0.15 < r/{\rm R}_{200} < 1.0$.  The inner boundary is imposed in order to avoid the region where baryons significantly affect the dark matter distribution, such that the NFW form generally does not provide a good fit.  We fit for the scale radius, $r_s$, and define the concentration in the usual way, as $c_{200, {\rm DM}} \equiv {\rm R}_{200}/r_s$.  In analogy to the dark matter concentration, we define the satellite concentration as $c_{200, {\rm sat}} \equiv {\rm R}_{200}/r_{s, {\rm sat}}$, where $r_{s, {\rm sat}}$ is the satellite scale radius, which we derive by fitting an NFW form to the cumulative satellite radial distributions shown in the top panel of Fig.~\ref{fig:rad_distrib}.  

Specifically, we treat the cumulative satellite radial distribution in analogy to the enclosed mass of a NFW profile whose value is equal to the total mass (or here total number of satellites) at ${\rm R}_{200}$ [i.e., ${\rm M}(r)/{\rm M}_{200} \rightarrow {\rm N} (d_{\rm sat})/{\rm N}_{\rm tot}({\rm R}_{200})$]:
\begin{equation}
\frac{{\rm N}(d_{\rm sat})}{{\rm N}_{\rm tot}({\rm R}_{200})} = \frac{f(x c_{200, {\rm sat}})}{f(c_{200, {\rm sat}})}   \ \ \ \ ,
\end{equation}

\noindent where
\begin{equation}
f(y) = \biggl[\ln{(1+y)} - \frac{y}{1+y}\biggr]    \ \ \ \ ,    
\end{equation}

\noindent and $x \equiv d_{\rm sat}/{\rm R}_{200}$, such that $x c_{200, {\rm sat}} = d_{\rm sat}/r_{s, {\rm sat}}$.

Note that, since the profile is required to pass through ${\rm N}(d_{\rm sat})/{\rm N}_{\rm tot}({\rm R}_{200})=1$ at $d_{\rm sat}={\rm R}_{200}$, which fixes the amplitude of the profile, the radial distribution depends only on the concentration parameter, $c_{200, {\rm sat}}$.  For each ARTEMIS halo and for the Milky Way and M31, we derive the best-fit concentration by simple chi-squared minimisation, adopting Poisson uncertainties on the cumulative radial profiles when fitting the model to the profiles.

The halo formation times are calculated as follows.  We first construct a mass growth history for the most massive progenitor of the host halo.  This is done by selecting all dark matter particles within ${\rm R}_{200}$ at $z=0$ and, using their unique IDs, identifying the friends-of-friends group with the largest fraction of these particles in previous snapshots (earlier times).  With the mass growth history, ${\rm M}_{200}(z)$, we identify the redshift where half of the final mass is in place \citep{lacey1993}.  We define the formation time, $t_{\rm form}$, as the lookback time to this redshift.

Our results suggest that the radial distribution of satellites does indeed `know about' the concentration of the host dark halo (and its formation time) at fixed halo mass, although the correlation is not particularly strong, with a Spearman coefficient of $\approx 0.3$. (We derive a similar correlation coefficient between the formation time of the halo and the satellite concentration.)  Nevertheless, there is a tendency for haloes that are more concentrated and/or that have formed early on, to also have more radially-concentrated satellite populations at present time.  

The filled, coloured points in Fig.~\ref{fig:c200} represented estimates of the concentrations for the Milky Way and M31.  The dark matter halo concentrations, $c_{200, {\rm DM}}$, for the Milky Way and M31 are from \citet{cautun2020} and \citet{tamm2012}, respectively.  Consistent with the findings from Fig.~\ref{fig:rad_distrib}, the Milky Way and M31 lie within the scatter of the simulations but do tend to be on the extremities of the simulation distribution.

\begin{figure}
\includegraphics[width=\columnwidth]{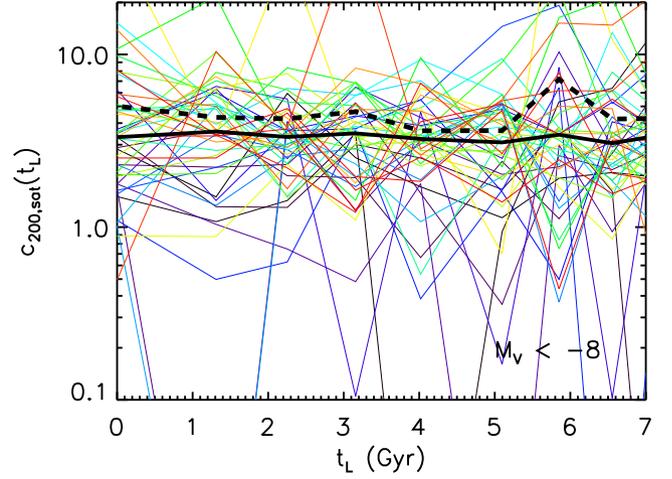}
\caption{The evolution of $c_{200, {\rm sat}}$ for individual ARTEMIS haloes as a function of lookback time, $t_L$.  Here we employ a selection criteria of $M_V < -8$ and that the satellites are within ${\rm R}_{200}(t_L)$.  The thick solid black curve represents the median concentration of the 45 ARTEMIS haloes as function of lookback time, while the thick dashed black curve corresponds to the mean trend.}
\label{fig:c200_evol}
\end{figure}

While the concentration of the satellite radial distribution appears to be partly set by that of the underlying dark matter, it is clear that this does not account for the bulk of the scatter in the satellite concentration at fixed halo mass (otherwise their would be a stronger correlation between the two concentration estimates).  Another possibility is that the true underlying radial distribution is relatively poorly sampled by the finite number of measured satellites used to trace it.  In other words, that the stochasticity in the positions of satellites along their orbits combined with the finite number of satellites used results in a noisy estimate of $c_{200, {\rm sat}}$.  We explore this possibility in Fig.~\ref{fig:c200_evol}, where we show the evolution of $c_{200, {\rm sat}}$ for individual haloes as a function of lookback time, $t_L$.  Here we employ a selection criteria of $M_V < -8$ and that the satellites are within ${\rm R}_{200}(t_L)$.  It is clear to see that, with this satellite selection criteria, an individual halo can vary significantly in its estimated concentration even on timescales of $\approx 1$ Gyr, which is about the dynamical timescale.  For a given halo, the mean RMS scatter about its mean concentration (averaged between $t_L=0$ and $t_L=7$ Gyr) is $\approx60\%$, such that a halo with $c_{200, {\rm sat}}=5$ can easily vary by $\pm3$ and with no significant growth in halo mass or abundance of satellites over that period.  

While the relatively weak correlation between $c_{200, {\rm sat}}$ and $c_{200, {\rm DM}}$ in Fig.~\ref{fig:c200} suggests that stochasticity plays a large factor in the satellite concentration, we note that the concentration of the host dark matter profile also evolves stochastically on relatively short timescales (see, e.g., \citealt{Wang2020}).  In analogy to Fig.~\ref{fig:c200_evol}, we have therefore examined the variation in the host concentration with time.  We find the mean RMS scatter about a given system's mean concentration to be $\approx20\%$ when the concentration is estimated from a profile including all dark matter particles, reducing to $\approx15\%$ when the contribution of satellites is removed from the host dark matter profile.  Thus, while the host concentration varies on relatively short timescales as well, these variations are not large enough to explain the bulk of the scatter in the satellite profile concentrations.

In summary, stochasticity is the most plausible explanation for the relatively large difference in the radial distributions of satellites of the Milky Way and M31 (i.e., the relatively small number of satellites used results in noisy estimates of the true radial profiles).  This stochasticity can be overcome by either increasing the number of satellites used to characterise the radial distribution (i.e., by including fainter satellites) or by using additional information about the orbits of satellites to measure, for example, their time-averaged radius rather than the instantaneous one.  By employing either (or both) of these measures, we might hope to better constrain the true underlying radial distribution and explore its link to the formation history of galaxies.

In Appendix \ref{sec:conc_def} we explore an alternative definition for the concentration of the satellite radial distribution, which is based on the radius enclosing half of the satellites (as opposed to the NFW scale radius).  While this quantity is more easily measured observationally, our conclusions above are unchanged with this alternative definition.

\subsubsection{Comparisons withe the Milky Way analogues outside the Local Group}
\label{sec:radial_LV_SAGA}

\begin{figure}
\includegraphics[width=\columnwidth]{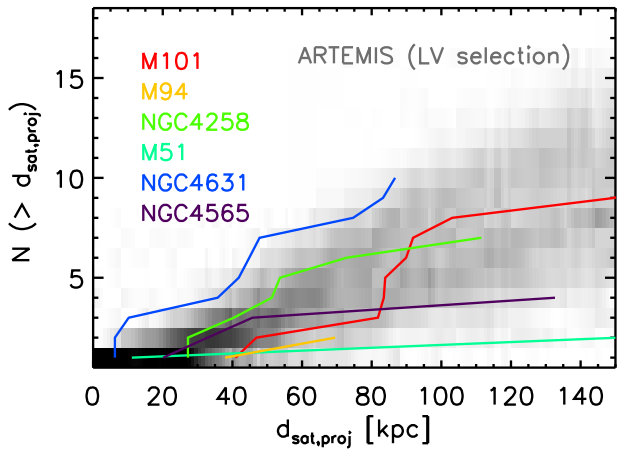}
\caption{The cumulative distributions of randomized projected distances, $d_{\rm sat, proj}$, of bright satellites ($M_{V} \le -9$, i.e the completeness magnitude of Local Volume observations) in the ARTEMIS galaxies, shown in grey-scale (heat map), in comparison with the observed distribution of projected distances of satellites of Milky Way analogues in the Local Volume, shown with various coloured lines. Only confirmed satellites are included for the observations.}
\label{fig:rad_proj_distrib_LV}
\end{figure}

Here we extend our comparison of radial distribution of simulated dwarfs with measurements in Milky Way analogues beyond the Local Group. The larger number of such systems allow us to gauge the intrinsic scatter in the observed distributions and compare it with the scatter obtained in the simulations. 

Fig.~\ref{fig:rad_proj_distrib_LV} shows the cumulative distributions of projected  distances of satellites in ARTEMIS compared with those of satellites in the six Milky Way-analogues from the Local Volume from \citet{carlsten2020_radial}. For the latter, we include only satellites that are distance-confirmed.  For the simulations, we compute projected distances from the 3D distances.  We randomly rotate the viewing angle through each system, drawing 1000 distributions per ARTEMIS halo and sum these to produce a `heat map' to show the resulting distribution.  We `column-normalise' the heat map (i.e., at a given project radius along the x axis we divide the pixels along the y axis by the sum of their values along that column) to better show the trend with projected distance.  Overall, there is reasonable agreement between simulations and observations, in the sense that the observed distributions are bracketed by the simulations.  Note that M51 exists on the very extremity of the simulated population, although it does have a few additional `possible satellites' (i..e, without distance confirmation) which, if confirmed, would bring M51 more in line with the simulations. 

\begin{figure}
\includegraphics[width=\columnwidth]{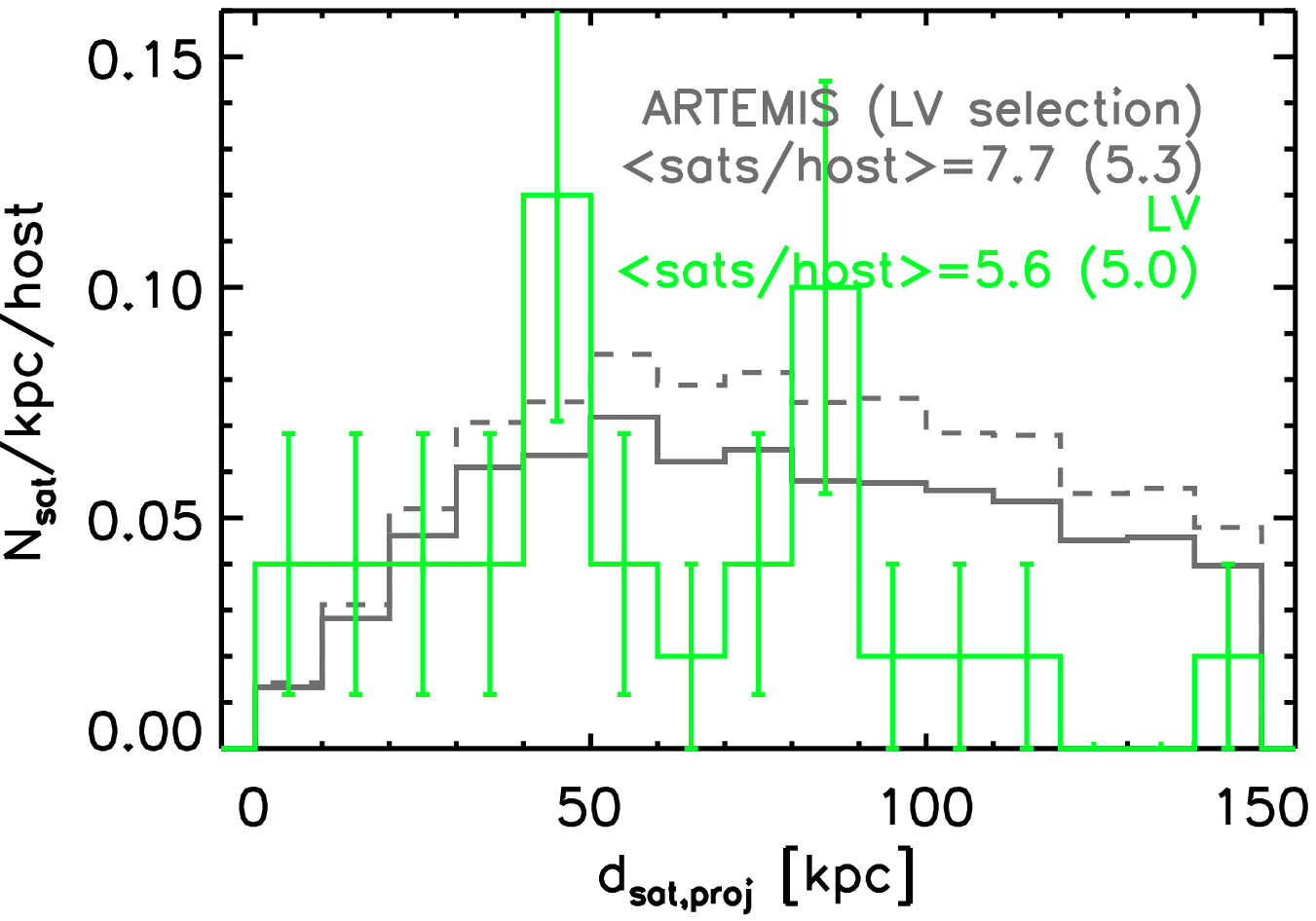} \\
\includegraphics[width=\columnwidth]{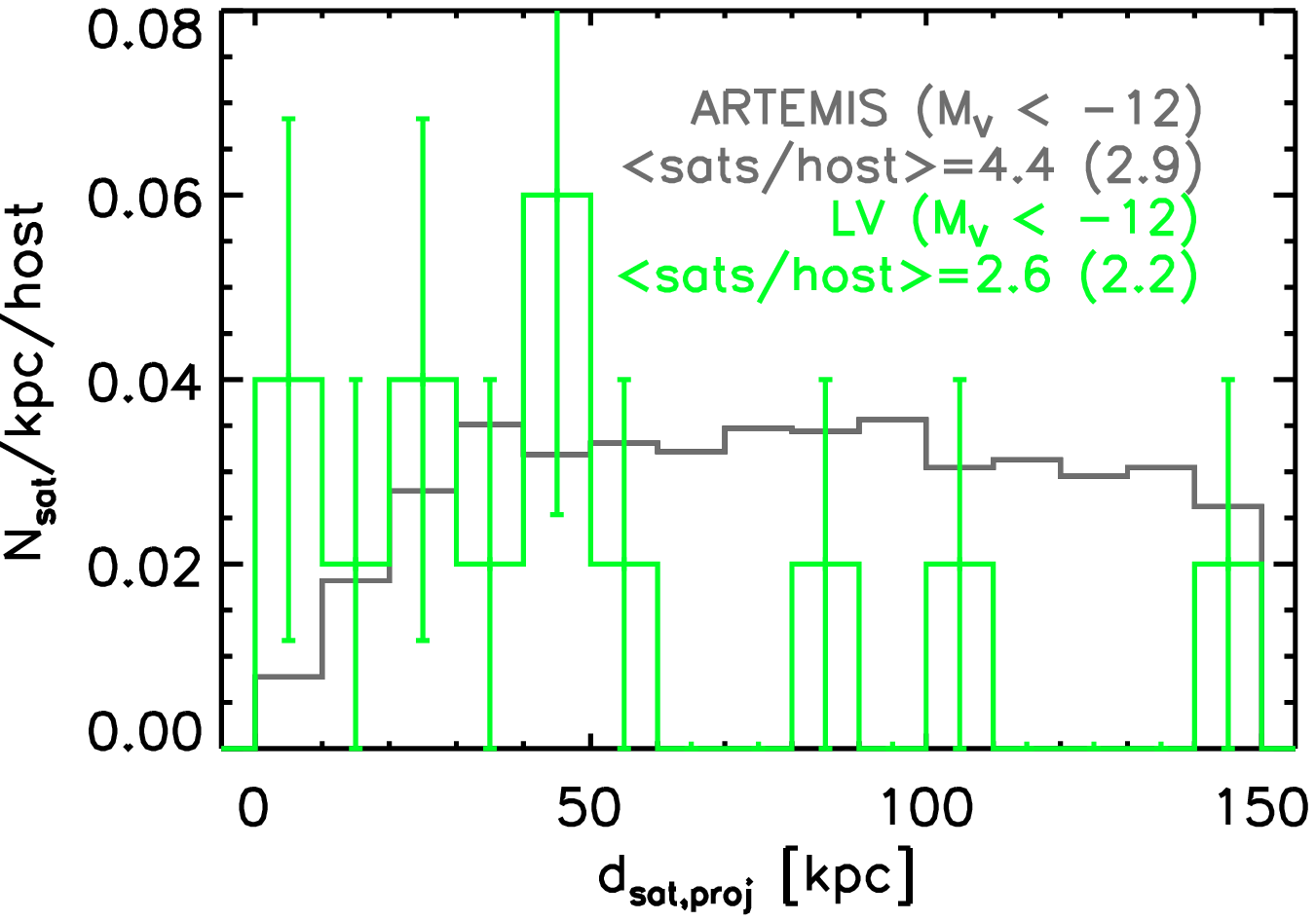}
\caption{{\it Top:} Normalized distributions of bright ($M_V<-9$) satellites (number per host per kpc) in Local Volume galaxies (green) and in ARTEMIS (black), versus $d_{\rm sat, proj}$. The dashed black histogram shows the impact of dropping the surface brightness limit when selecting satellites from the simulations.  {\it Bottom:} As above, but for satellites brighter than $M_V<-12$. The numbers in the legend give the mean number of satellites contained with 150 kpc (100 kpc).}
\label{fig:hist_rad_proj_distrib_LV}
\end{figure}

Fig.~\ref{fig:hist_rad_proj_distrib_LV} provides a more quantitative analysis of these radial distributions.  Following \citet{carlsten2020_radial}, we plot the normalized differential distributions of satellites (number per host per kpc) in the six Local Volume galaxies (with green) and in ARTEMIS (with black lines), versus satellite projected distances. The top panel uses the default LV selection criteria for the simulations (i.e., $M_V \le -9$, $\mu_{{\rm eff},V} \le 28.3$ mags. arcsec$^{-2}$ and $d_{\rm sat,proj} \le 150$ kpc).  The comparison is somewhat hindered by the sparsity of Local Volume data.  Note that error bars on the Local Volume data correspond to Poisson errors.  The simulations, on the other hand, include $45$ hosts where the satellites are viewed from a thousand random angles each. Nevertheless, this analysis shows that cosmological models are in reasonably good agreement with the observations in the top panel of Fig.~\ref{fig:hist_rad_proj_distrib_LV}.  There is a tendency for the simulations to predict a somewhat higher abundance of satellites relative to the observations at larger radii ($\ga100$~kpc), but this is likely due to the Local Volume observations not fully extending to $150$~kpc for all hosts in that sample.  In the legend in the top panel of Fig.~\ref{fig:hist_rad_proj_distrib_LV} we provide the mean number of satellites contained within $150$~kpc and $100$~kpc (the latter in parenthesis) for the Local Volume and ARTEMIS samples.  There is reasonably good agreement within 100 kpc.  For reference, the dashed black histogram shows the impact of dropping the surface brightness limit when selecting satellites from the simulations.  Overall the effect is fairly modest, typically increasing the satellite abundance by 10-15\% with a mild dependence on projected distance.

\begin{figure}
\includegraphics[width=\columnwidth]{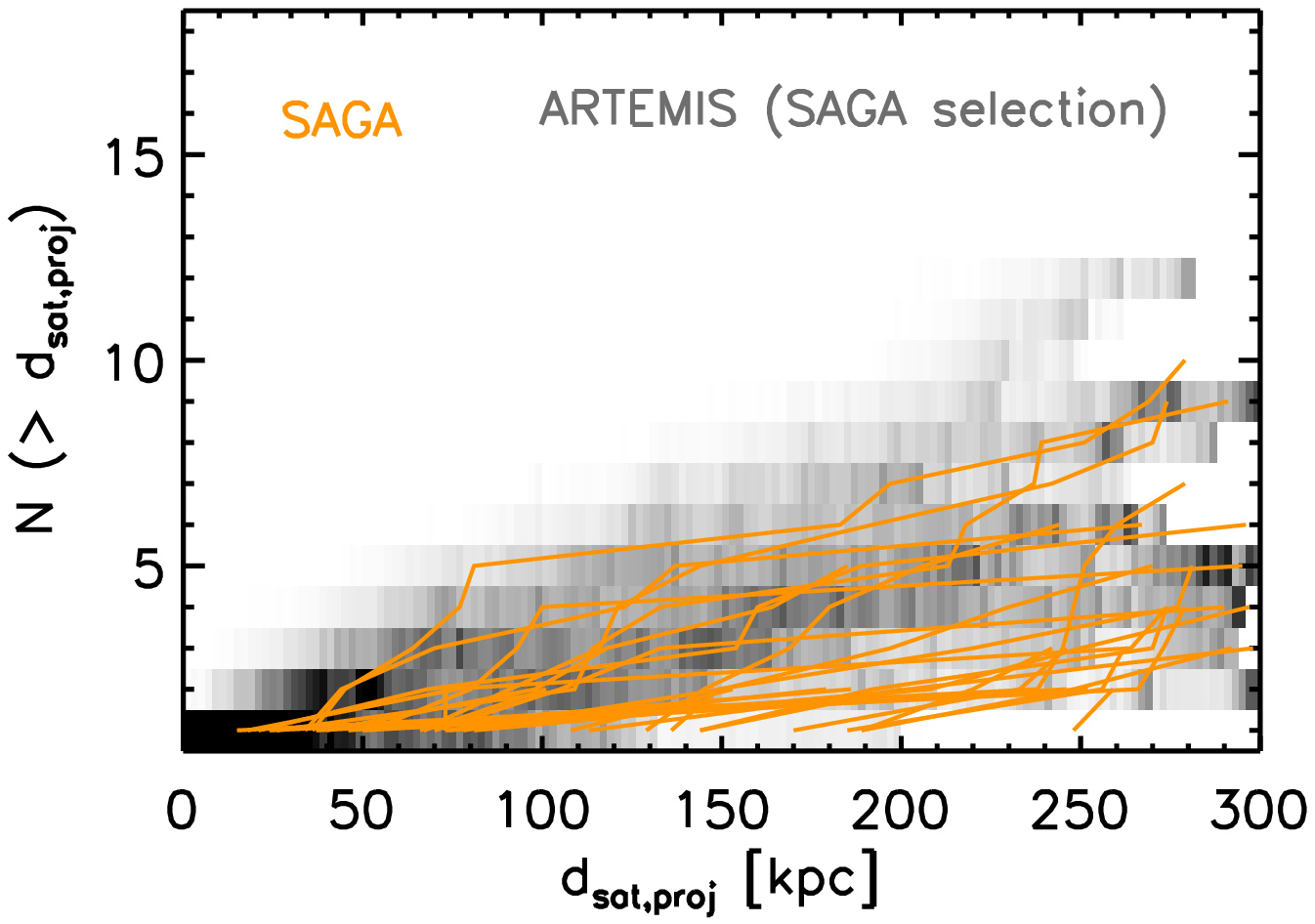}
\includegraphics[width=\columnwidth]{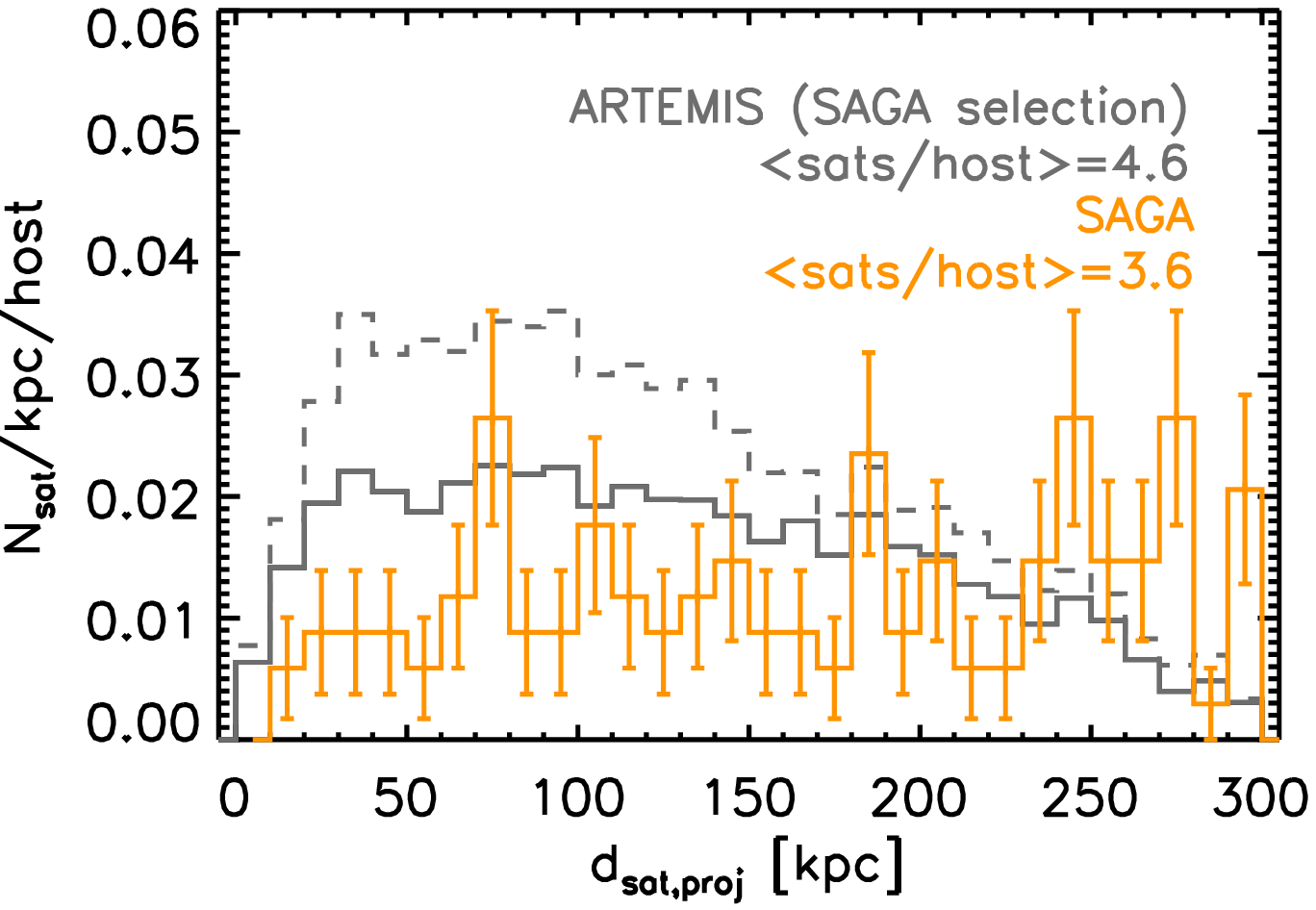}
\includegraphics[width=\columnwidth]{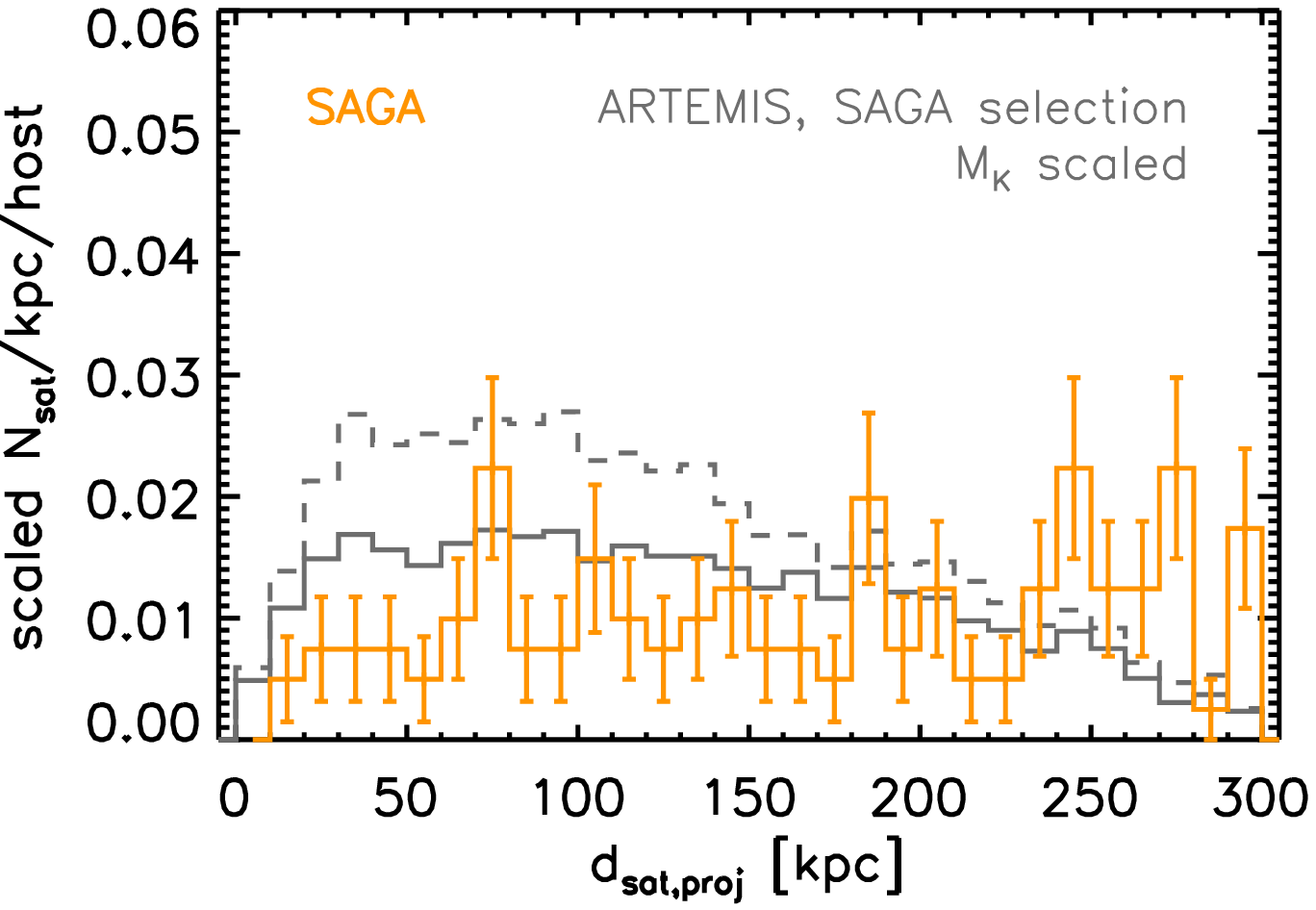}
\caption{{\it Top:} The cumulative distributions of randomized projected distances, $d_{\rm sat, proj}$, of ARTEMIS satellites applying the SAGA selection criteria}, shown as a column-normalised heat map, in comparison with the observed distribution of projected distances of satellites of Milky Way analogues in SAGA, shown with orange lines. {\it Middle:} Normalized differential distribution functions of satellites per kpc per host in ARTEMIS compared with SAGA.  The solid black histogram represents the simulations when the full SAGA selection criteria is applied, while the dashed black histogram shows the result when the surface brightness cut is removed.  {\it Bottom:} Same as the middle panel except that the histograms have been scaled in amplitude according to eqn.~\ref{eq:nsat_mk_saga_sim} in order to account for the difference in the mean host K-band magnitude of ARTEMIS and SAGA (see text).
\label{fig:rad_proj_distrib_SAGA}
\end{figure}

In the bottom panel of Fig.~\ref{fig:hist_rad_proj_distrib_LV} we restrict the comparison to only the brightest satellites ($M_V <-12$). As discussed in \citet{carlsten2020_radial}, the Local Volume satellite populations appear to be more radially concentrated in the case of the brightest ($M_V <-12$) dwarfs compared with those of the entire sample ($M_V <-9$). On the other hand, the simulations obtain similar distributions independent of the satellite magnitude cut.  In terms of the difference with the Local Volume observations, we caution that there are very large Poisson uncertainties associated with the observed distribution, as there are typically only 2 to 3 satellites per halo meeting this selection criteria, derived from a sample of only 6 hosts.  Furthermore, as already noted, the full area within 150 kpc was not surveyed for all 6 hosts.  

In Fig.~\ref{fig:rad_proj_distrib_SAGA} we show the cumulative (top panel) and differential (middle panel) projected radial distributions of satellites in ARTEMIS with the SAGA selection criteria applied and make comparisons with the respective distributions from SAGA (orange curves).  Overall the agreement with SAGA is reasonably good.  There appears to be a slight difference in the radial dependence, with ARTEMIS having a stronger contribution from smaller projected radii (though perhaps not as strong as in the Local Volume results presented above).

The dashed-black histogram in the middle panel of Fig.~\ref{fig:rad_proj_distrib_SAGA} shows the impact of dropping the surface brightness cut on the predicted differential radial distribution from ARTEMIS.  The effect is rather large, particularly at distances of $\la 150$ kpc, where the abundance of satellites can increase by more than 50\% as a result of dropping the surface brightness limit.  The relatively strong radial dependence of this effect suggests that environmental processes (e.g., tidal heating) are driving it, but we leave a detailed analysis of the causes of this effect for future work.

Finally, in the bottom panel of Fig.~\ref{fig:rad_proj_distrib_SAGA} we show the impact of scaling the mean differential curves in the middle panel by a correction factor based on eqn.~\ref{eq:nsat_mk_saga_sim} that accounts for differences in the mean host K-band magnitudes of ARTEMIS and SAGA.  As the mean host K-band magnitude of ARTEMIS is larger (brighter) than for SAGA (both of which are brighter than the adopted pivot point of $-23.5$), the correction factor is larger for ARTEMIS than SAGA, resulting in a slightly improved agreement in the amplitude of the predicted and observed radial distributions.

\section{Conclusions}
\label{sec:concl}

The advent of deep observational surveys of local Milky Way `analogues' and their satellite populations allows us to place the Milky Way in a broader cosmological context and to test the realism of cosmological simulations.  In the present study, we have used the ARTEMIS suite of cosmological hydrodynamical simulations, introduced recently in \citet{font2020}, to make comparisons with the satellite luminosity functions, radial distribution functions, and abundance scaling relations of satellite galaxies in the Milky Way, M31, and Milky Way analogues sampled in the Local Volume \citep{carlsten2020_survey,carlsten2020_radial,carlsten2021} and SAGA \citep{geha2017,mao2021} surveys.  The main findings of these comparisons are as follows:

\begin{itemize}
    \item An analysis of the magnitude--surface brightness relations of the SAGA and Local Volume surveys indicates that in order to enable a fair comparison with the simulations, both magnitude and surface brightness limits need to be factored in (Fig.~\ref{fig:mueff_mag}).
    \item The distribution of {\it host} $K$-band luminosities, a good proxy for stellar mass, of ARTEMIS galaxies is very similar to that of SAGA and Local Volume samples (Fig.~\ref{fig:hist_mag}), ensuring a fair comparison of their satellite populations.  Nevertheless, we also explored the impact of  differences in $K$-band magnitudes throughout. 
    \item The simulated satellite luminosity functions for the 45 ARTEMIS haloes are very similar to that observed for the Milky Way and M31 (Fig.~\ref{fig:lfunc_MW_M31}) and for the Local Volume and SAGA surveys (Fig.~\ref{fig:lfunc_sat}) even though no aspect of the simulations was adjusted to obtain this.  Furthermore, the simulations appear to naturally capture the large halo-to-halo diversity in the shape and amplitudes of the luminosity functions, having systems as abundant as M31 and as sparse as M94.
    \item \citet{carlsten2021} and \citet{mao2021} have reported evidence for a slight excess of bright ($M_V \la -18$) satellites with respect to previous simulations.  In contrast, we find that ARTEMIS reproduces the abundance of bright satellites in the SAGA and Local Volume surveys to within Poisson uncertainties (Fig.~\ref{fig:lfunc_diff}).  
    \item Contrary to previous claims, we find $\Lambda$CDM-based simulations have no difficulties in reproducing the large magnitude gaps present in some observed satellite luminosity functions (Fig.~\ref{fig:sat_gap}).
    \item The abundance of satellites depends strongly on host properties, including stellar mass and (especially) total halo mass.  However, applying practical observational selection criteria, such as fixed bright magnitude limits and fixed physical apertures, reduces the strength of these correlations (Fig.~\ref{fig:Nsat_scaling}).  We provide power law fits to the various scaling relations in eqns.~\ref{eq:nsat_mstar}--\ref{eq:nsat_mk_saga_sim}.  Contrary to some recent claims, we find no significant correlation between satellite abundance and host morphology (at fixed halo mass) in either the simulations or observations (Fig.~\ref{fig:Nsat_BT}).
    \item The radial distribution of satellites in the simulations is compatible with that observed for the Milky Way and M31 (Fig.~\ref{fig:rad_distrib}).  The Milky Way has a considerably more concentrated radial profile than M31. While this could potentially be explained if Milky Way's underlying dark matter halo was significantly more concentrated than that of M31, recent measurements suggest that it is M31 that has the higher dark matter concentration (Fig.~\ref{fig:c200}).  It is more likely that stochasticity due to the use of a finite number of satellites is the main cause of the difference between M31 and the Milky Way.  We have shown that the inferred radial concentration of the satellite population a given halo can vary significantly on a dynamical timescale when imposing observational selection criteria (Fig.~\ref{fig:c200_evol}).
    \item The simulated radial distributions are also compatible with those measured for Milky Way analogues in the Local Volume (Figs.~\ref{fig:rad_proj_distrib_LV} and \ref{fig:hist_rad_proj_distrib_LV}) and SAGA (Fig.~\ref{fig:rad_proj_distrib_SAGA}) surveys, so long as the appropriate observational selection criteria are applied.
\end{itemize}

The present study has focused mainly on the luminosity functions and radial distribution functions of satellites around Milky Way-mass haloes.  Broadly speaking, the simulations reproduce the observed distributions and with no fine tuning to do so.  To yield sensible satellite populations in this regard requires not only having a realistic `backbone' for structure formation (which $\Lambda$CDM appears to provide) but also that processes such as star formation and stellar feedback are at least reasonable, otherwise the simulations would populate the dark matter haloes with galaxies of incorrect stellar mass/luminosity.  However, a possibly much more challenging test for the simulations will be whether they can also reproduce the diversity of internal properties of satellites, including their star formation rates, colours, gas fractions, chemical abundances, and so on, and correctly describe how these quantities depend on host properties, e.g., halo mass, concentration, etc.  We plan to examine these questions in future work.  

\section*{Acknowledgments}
The authors would like to thank the referee for constructive comments that substantially improved the paper.
The authors would also like to thank the members of the EAGLE team for making their cosmological simulation code available for the ARTEMIS project.  They also thank the Local Volume and SAGA survey teams for making their data sets publicly available.  The authors would also like to thank Scott Carlsten for providing data and comments on the draft and Paul Bennet and Denija Crnojevi{\'c} for useful discussions. This project has received funding from the European Research Council (ERC) under the European Union's Horizon 2020 research and innovation programme (grant agreement No 769130). This work used the DiRAC@Durham facility managed by the Institute for Computational Cosmology on behalf of the STFC DiRAC HPC Facility. The equipment was funded by BEIS capital funding via STFC capital grants ST/P002293/1, ST/R002371/1 and ST/S002502/1, Durham University and STFC operations grant ST/R000832/1. DiRAC is part of the National e-Infrastructure.

\section*{Data availability}
The data underlying this article may be shared on reasonable request to the corresponding author.

%%%%%%%%%%%%%%%%%%%%%%%%%%%%%%%%%%%%%%%%%%%%%%%%%%



\bibliographystyle{mnras}
\bibliography{references} % if your bibtex file is called example.bib
%%%%%%%%%%%%%%%%%%%%%%%%%%%%%%%%%%%%%%%%%%%%%%%%%%

%%%%%%%%%%%%%%%%% APPENDICES %%%%%%%%%%%%%%%%%%%%%
\appendix
\section{ARTEMIS host properties}
\label{sec:appendixA}

In Table \ref{tab:table1} we list various properties of the ARTEMIS host haloes and central galaxies.

\begin{table*}
	\centering
	\caption{The main properties of Milky Way-analog haloes in the ARTEMIS simulations. The columns include: the ID name of the simulated galaxy, the ${\rm M}_{200}$ mass, the virial mass (${\rm M}{\rm vir}$), the ${\rm R}_{200}$, virial radius (${\rm R}_{\rm vir}$), and the magnitudes $M_B$, $M_V$, $M_K$ and $M_r$ for each galaxy.}
	\label{tab:table1}
	\begin{tabular}{crrcccccc} % four columns, alignment for each
		\hline
		Galaxy & ${\rm M}_{200}$ &  ${\rm M}_{\rm vir}$ & ${\rm R}_{200}$ & ${\rm R}_{\rm vir}$ & $M_{B}$ & $M_{V}$ & $M_{K}$ & $M_{r}$\\
                & $[10^{12} \,{\rm M}_{\odot}]$ &  $[10^{12} \,{\rm M}_{\odot}]$ &  [kpc] & [kpc] & [mag] & [mag] & [mag] & [mag] \\
		\hline
   G1  & 1.19   &  1.36   &  218.59 &   284.27  & -21.22 &  -21.69 &  -24.05 &   -21.86\\
 G2  &    1.65    &  1.90   &  243.89 &  317.47  & -21.37  & -21.94 &  -24.32 &  -22.14\\
 G3  &    1.70     &  1.96     & 246.25   & 320.55    &-20.57  &  -21.29 &   -24.01    &-21.56\\
 G4  &    1.43      & 1.64      &232.49   & 302.41  &  -19.96  &  -20.82 &   -23.68   & -21.12\\
 G5  &    1.64      & 1.89    &   243.36  &  316.78   & -20.41   & -21.17 &   -23.91  &  -21.44\\
 G6  &    1.64    &   1.89     & 243.47   & 316.93   & -21.14   & -21.80 &   -24.45  &  -22.05\\
 G7  &    0.99    &  1.14     & 206.05   & 267.82  &  -20.34    &-20.97 &   -23.53   & -21.20\\
 G8  &    1.63    &   1.87   &   242.80   & 316.05  &  -20.42   & -20.92 &   -23.42   & -21.12\\
 G9  &    1.11    &   1.27     & 213.53   & 277.57  &  -21.35   & -21.83 &   -24.16  &  -22.01\\
 G10  &    1.15     &  1.32     & 216.27  &  281.12   & -19.40    &-20.28 &   -23.19   & -20.60\\
 G11  &    1.17      & 1.34     & 217.28  &  282.45  &  -20.42   & -21.23 &   -24.08   & -21.53\\
 G12  &    1.32     &  1.52    &  226.48  &  294.60  &  -21.07  &  -21.68 &   -24.20   & -21.90\\
 G13  &    1.17     &  1.34    &  217.28   & 282.45   & -20.64   & -21.16 &   -23.54   & -21.35\\
 G14  &    1.22    &   1.39     & 220.14  &  286.36   & -20.04    &-20.79 &   -23.60    &-21.06\\
 G15  &    1.12     &  1.28     & 214.32   & 278.60   & -21.31  &  -21.83 &   -24.18  &  -22.02\\
 G16  &    1.27     &   1.45     & 223.31   & 290.47   & -21.07  &  -21.57  &  -23.94   & -21.75\\
 G17  &    1.17     &  1.34    &  217.28  &  282.44  &  -21.22  &  -21.70 &   -24.05  &  -21.88\\
 G18  &    0.97     & 1.11     & 204.03   & 265.19  &  -20.76  &  -21.26  &  -23.69  &  -21.45\\
 G19  &    0.96   &   1.10    &  203.66   & 264.72  &  -20.39  &  -20.97  &   -23.49  &  -21.18\\
 G20  &    1.06     &  1.21    &  210.16  &  273.17  &  -20.69  &  -21.32  &   -23.91  &  -21.56\\
 G21  &    1.01      & 1.16     & 207.06  &  269.14  &  -18.94   & -19.84  &  -22.72  &  -20.16\\
 G22  &    1.01     &  1.15    &   206.84  &   268.85  &   -20.07  &   -20.77  &  -23.50  &  -21.03\\
 G23  &    0.99    &  1.14    &  205.92   & 267.66  &  -20.51  &   -21.09  &  -23.66  &  -21.31\\
 G24  &    1.03      & 1.18    &  208.27  &  270.72   & -21.14    &-21.64 &   -24.05  &  -21.83\\
 G25  &    0.91    &  1.04    &  200.02  &  259.81   & -20.92    &-21.40 &   -23.76  &  -21.59\\
 G26  &    0.89    &  1.02    &  198.86   & 258.29  &  -20.76  &  -21.32  &  -23.89  &  -21.53\\
 G27  &    0.79   &   0.91   &  191.18   & 248.32  &  -20.94  &  -21.43 &   -23.77  &  -21.62\\
 G28  &    0.76    &  0.87   &  188.83  &  245.27   & -20.83  &  -21.32  &  -23.66  &  -21.50\\
 G29  &    0.88    &  1.01    &  197.87   & 257.03   & -20.71    &-21.22 &   -23.73  &  -21.42\\
 G30  &    0.81   &   0.92   &  192.16 &   249.60  &  -20.38  &  -20.99 &   -23.57   & -21.22\\
 G31  &    0.83   &   0.95   &  193.99   & 251.98   & -20.80  &  -21.21 &   -23.50   & -21.37\\
 G32  &    0.78   &   0.89   &  190.51  &  247.45   & -20.63  &  -21.15  &  -23.58  &  -21.34\\
 G33  &    0.78   &   0.89    & 189.92   & 246.68    &-20.12    &-20.82 &   -23.51   & -21.08\\
 G34  &    0.79    &  0.90    & 190.65  &  247.62  &  -20.60   & -21.14 &   -23.61  &  -21.34\\
 G35  &    0.68    &  0.78    & 181.57  &  235.68  &  -19.97  &  -20.59 &   -23.15  &  -20.83\\
 G36  &    3.64    &   4.21     & 317.18   & 413.78   & -20.24   & -21.11  &   -24.00  &  -21.42\\
 G37  &    0.66   &   0.76   &  180.11   & 233.78   & -18.89   & -19.81  &   -22.76  &  -20.13\\
 G38  &    0.71   &   0.81   &   184.35  &  239.28  &  -20.65   & -21.22   & -23.70  &   -21.43\\
 G39  &    0.75    &  0.85   &  187.25  &  243.17  &  -20.04   & -20.67  &   -23.21  &  -20.91\\
 G40  &    0.76   &   0.86   &   187.99  &   244.18  &   -20.71   & -21.17  &  -23.45  &  -21.33\\
 G41  &    0.69   &   0.78   &   182.18   & 236.47  &   -19.27  &  -20.12  &  -22.99  &   -20.43\\
 G42  &    0.72    &  0.82   &  184.68  &   239.71  &   -20.11  &  -20.69  &   -23.29  &  -20.92\\
    \hline
G43 & 1.97  & 2.27  & 258.79  & 337.10  & -22.45  & -22.83  & -25.00  & -22.97\\    
G44 & 1.62  & 1.86  & 235.20  & 306.16  & -20.87  & -21.53 & -24.14  & -21.78\\    
G45 & 1.37  & 1.57  &  222.76 & 289.77  & -21.90  & -22.43  & -24.79  & -22.62\\
    \hline
	\end{tabular}
\end{table*}

\section{Alternative concentration definition}
\label{sec:conc_def}

In Section \ref{sec:radial_MW_M31} we characterised the satellite radial distributions of the ARTEMIS haloes and the Milky Way and M31 via a concentration parameter derived by fitting an NFW profile to the radial distributions.  Here we explore an alternative definition of concentration, which is potentially more easily measured for systems with few satellites.  Specifically, here we define the satellite concentration as $c_{1/2, {\rm sat}} \equiv {\rm R}_{200}/{\rm R}_{1/2}$, where ${\rm R}_{1/2}$ is the radius enclosing 50\% of the satellites within ${\rm R}_{200}$.

\begin{figure}
\includegraphics[width=\columnwidth]{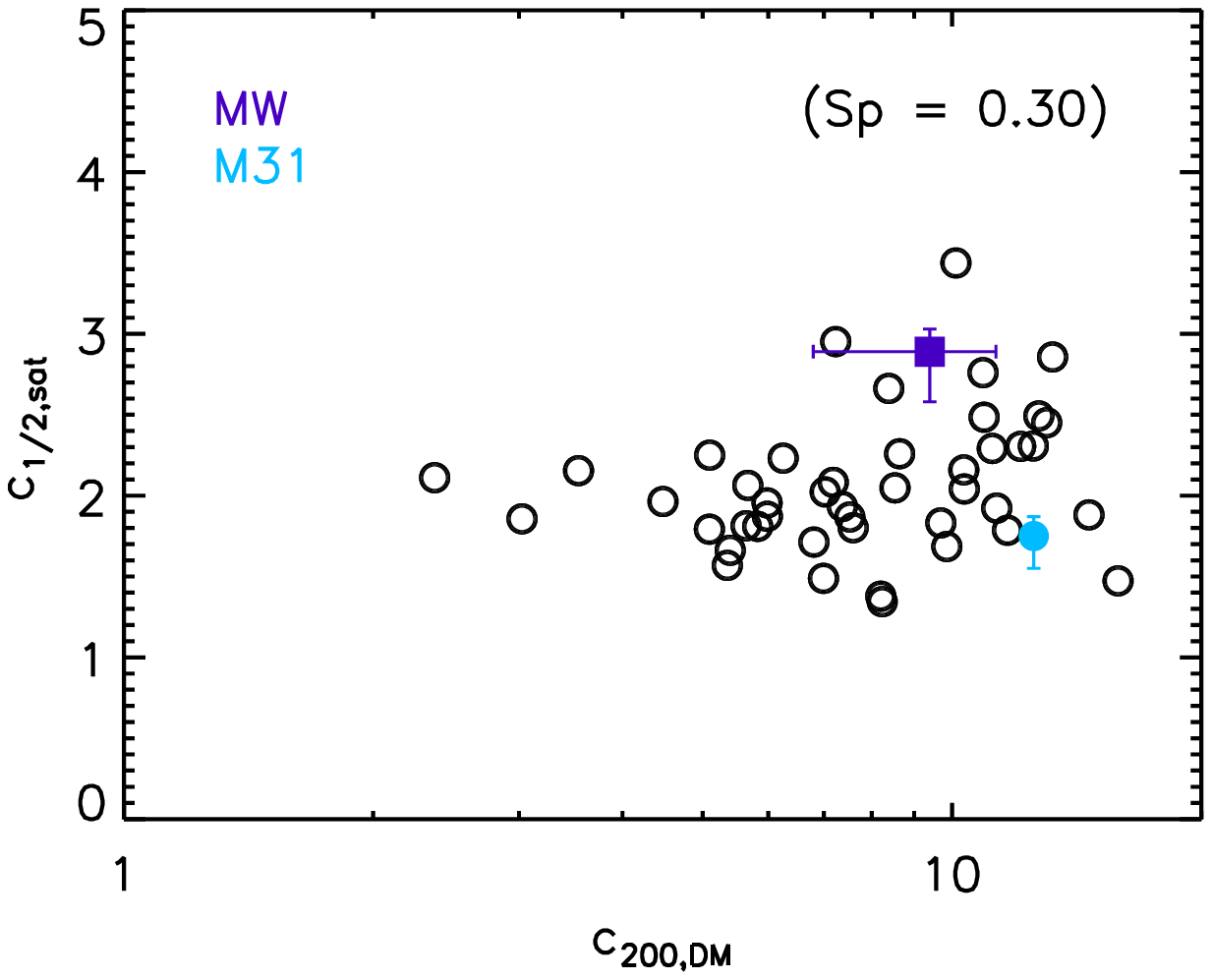}
\caption{The relation between the concentration of the satellite population, $c_{1/2, {\rm sat}} \equiv {\rm R}_{200}/{\rm R}_{1/2}$ (where ${\rm R}_{1/2}$ is the radius enclosing $50\%$ of satellites, in units of ${\rm R}_{200}$) and the concentration of dark matter halo of the simulated host galaxies, $c_{200, {\rm DM}}$.  The correlation is relatively weak (Spearman coefficient of $\approx 0.3$), but there is a tendency for haloes that are highly concentrated (and that formed earlier) to also have more radially concentrated satellites.}
\label{fig:chalf}
\end{figure}

\begin{figure}
\includegraphics[width=\columnwidth]{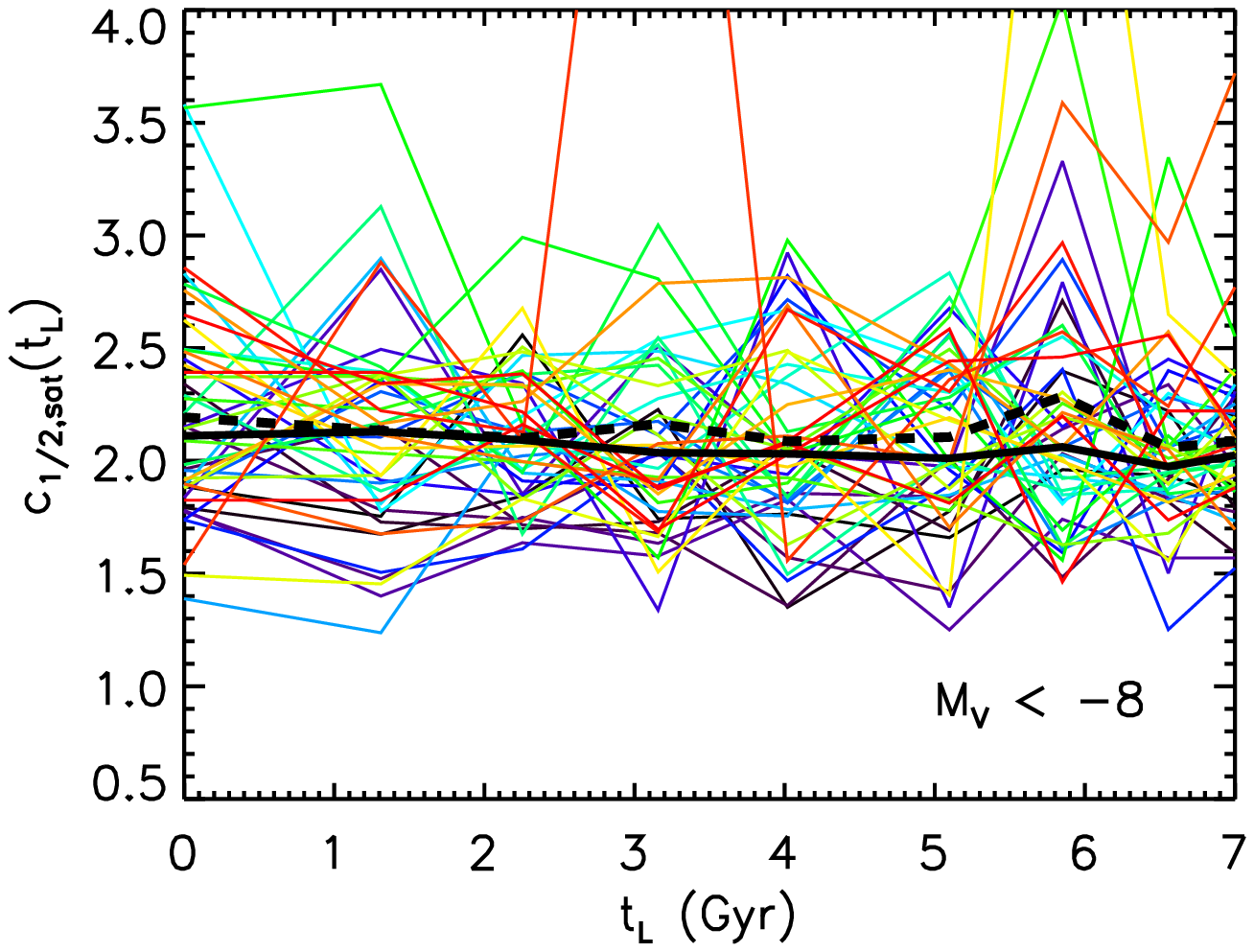}
\caption{The evolution of $c_{{1/2}, {\rm sat}}$ for individual ARTEMIS haloes as a function of lookback time, $t_L$.  Here we employ a selection criteria of $M_V < -8$ and that the satellites are within ${\rm R}_{200}(t_L)$.  The thick solid black curve represents the median concentration of the 45 ARTEMIS haloes as function of lookback time, while the thick dashed black curve corresponds to the mean trend.}
\label{fig:chalf_evol}
\end{figure}

Indeed, as shown in Fig.~\ref{fig:chalf}, we find that this definition of concentration has a reduced scatter compared to the NFW-based definition (20\% here as opposed to 60\% for the NFW concentration).  Nevertheless, our conclusions remain the same: the correlation between this alternative concentration and the dark matter halo concentration is present but weak and stochasticity (see Fig.~\ref{fig:chalf_evol}) is the main cause of scatter in this concentration at fixed halo mass.  Also in agreement with the results presented in the main text, we find that the Milky Way and M31 have concentrations consistent with the simulated population.

Note that, although a concentration based on ${\rm R}_{1/2}$ is perhaps easier to measure (less noisy) than the NFW-based concentration, it is also less likely to be correlated with the formation history of a halo, as it is less sensitive to the inner regions which collapsed earlier.  In other words, it has a smaller `lever arm' with respect to the formation history compared to the NFW-based concentration.

% Don't change these lines
\bsp	% typesetting comment
\label{lastpage}
\end{document}